\documentclass[twocolumn,astrosymb,tighten]{aastex631}
\usepackage{xcolor}
\usepackage{amsmath}
\usepackage{graphicx}
\usepackage{tabularx}

\usepackage{float}
\usepackage[mathscr]{euscript}

\newcommand{\revision}[1]{#1}

\begin{document}

\title{Magnetic field alignment with dense cores in the transition between cloud and core scales}

\correspondingauthor{Sean Yin}

\author[0009-0004-8142-7897]{Sean Yin}
\affiliation{David A. Dunlap Department of Astronomy and Astrophysics, University of Toronto, 50 St. George St, Toronto, ON M5S 3H4, Canada}
\affiliation{Department of Physics, Engineering Physics \& Astronomy, Queen’s University, 64 Bader Lane, Kingston, ON K7L 3N6, Canada}
\email{sean.yin@queensu.ca}

\author[0000-0002-8897-1973]{Ayush Pandhi}
\affiliation{\revision{Department of Physics, McGill University, 3600 rue University, Montr\'eal, QC H3A 2T8, Canada}}
\affiliation{\revision{Trottier Space Institute, McGill University, 3550 rue University, Montr\'eal, QC H3A 2A7, Canada}}
\affiliation{David A. Dunlap Department of Astronomy and Astrophysics, University of Toronto, 50 St. George St, Toronto, ON M5S 3H4, Canada}
\affiliation{Dunlap Institute for Astronomy and Astrophysics, University of Toronto, 50 St. George Street, Toronto, ON M5S 3H4, Canada}

\author[0000-0001-7594-8128]{Rachel Friesen}
\affiliation{David A. Dunlap Department of Astronomy and Astrophysics, University of Toronto, 50 St. George St, Toronto, ON M5S 3H4, Canada}

\author[0000-0002-0859-0805]{Simon Coud\'e}
\affiliation{\revision{National Astronomical Observatory of Japan, National Institute of Natural Sciences, 2-21-1 Osawa, Mitaka, Tokyo 181-8588, Japan}}
\affiliation{Department of Earth, Environment, and Physics, Worcester State University, Worcester, MA 01602, USA}
\affiliation{Center for Astrophysics $\vert$ Harvard \& Smithsonian, 60 Garden Street, Cambridge, MA 02138, USA}

\author[0000-0002-4666-609X]{Laura Fissel}
\affiliation{Department of Physics, Engineering Physics \& Astronomy, Queen’s University, 64 Bader Lane, Kingston, ON K7L 3N6, Canada}

\author[0000-0001-7474-6874]{Sarah Sadavoy}
\affiliation{Department of Physics, Engineering Physics \& Astronomy, Queen’s University, 64 Bader Lane, Kingston, ON K7L 3N6, Canada}

\author[0000-0002-9289-2450]{James Di Francesco}
\affiliation{NRC Herzberg Astronomy and Astrophysics, 5071 West Saanich Road, Victoria, BC V9E 2E7, Canada}
\affiliation{Department of Physics and Astronomy, University of Victoria, Victoria, BC V8P 5C2, Canada}

\author[0000-0002-6773-459X]{Doug Johnstone}
\affiliation{NRC Herzberg Astronomy and Astrophysics, 5071 West Saanich Road, Victoria, BC V9E 2E7, Canada}
\affiliation{Department of Physics and Astronomy, University of Victoria, Victoria, BC V8P 5C2, Canada}

\author[0000-0002-5391-5568]{Fr\'ed\'erick Poidevin}
\affiliation{Instituto de Astrof\'{\i}sica de Canarias, V\'{\i}a  L\'actea, 38205 La Laguna, Tenerife, Spain}
\affiliation{Universidad de La Laguna, Departamento de Astrof\'{\i}sica, 38206 La Laguna, Tenerife, Spain}

\author[0000-0001-8749-1436]{Mehrnoosh Tahani}
\affiliation{Department of Physics \& Astronomy, University of South Carolina, Columbia, SC 29208, USA}
\affiliation{Kavli Institute for Particle Astrophysics \& Cosmology (KIPAC), Stanford University, Stanford, CA 94305, USA}

\begin{abstract}
In a magnetically-dominated model of star formation, we expect to see alignments between the magnetic field orientation of star-forming dense cores and the cloud-scale magnetic field. \cite{Pandhi2023} showed instead, however, that the orientation of cores and their angular momentum vectors appear random with respect to the larger-scale magnetic field, implying that magnetic fields may play a diminished role in core formation and evolution. Here, we use higher-resolution dust polarization data from the B-Fields In Star-forming Region Observations (BISTRO) survey on the James Clerk Maxwell Telescope (JCMT) to investigate the change in the magnetic field orientation from cloud scales to core scales, and reassess any correlations between core-scale magnetic fields, core orientations and core velocity gradients. We produce a catalog of 79 cores over 14 star-forming regions with averaged core-scale magnetic field orientations. We find that the core-scale magnetic field is more disordered compared to the cloud-scale field, as measured by an increased standard deviation in the magnetic field vector orientations. Alignment between the core-scale and cloud-scale field varies greatly between regions. Our results are consistent with random alignments between the core-scale magnetic field, core orientation, and core velocity gradient, in agreement with the results by \cite{Pandhi2023} for the cloud-scale field. We conclude that there is a clear change in the magnetic field in the transition from cloud- to core-scales. Our results suggest that the magnetic field may not play a dominant role in the evolution of dense cores on core scales.

\end{abstract}

\section{Introduction} \label{sec:intro}

Stars form from local overdensities within molecular clouds known as dense cores \citep{1983_Myers}. The dynamics of these dense cores are mainly driven by a combination of gravity, turbulence, and magnetic fields \citep[][and references therein]{Ballesteros-Paredes2007}. If magnetic fields dominate the dynamics of dense cores, we expect to see correlations between core properties and the magnetic field \citep{Pattle2023}. Dense cores that are most likely to be unstable to gravitational collapse are primarily embedded in filaments within molecular clouds. Observationally, filamentary gas structures are preferentially aligned with the cloud-scale magnetic field at low column density, and they transition to a perpendicular alignment at higher column densities \citep{2016Planck}. Some studies show additional transitions in magnetic field orientation along filaments \citep[e.g.,][]{pillai_2020}. At the higher densities found in star-forming cores, the magnetic field strength $B$ increases with number density $n$ following $B \propto n^{0.65}$ \citep{2010_Crutcher}. On these scales, individual cores show a variety of magnetic field morphologies from pinched or hourglass shapes \cite[e.g.,][]{2006Girart, 2014Qiu} to ordered fields that are aligned or misaligned with the large-scale cloud field \cite[e.g.,][]{BISTRO_OphC, BISTRO_L1689}.

This variation in the magnetic field morphology on core scales \revision{may} impact significantly the subsequent mass accretion from cloud to core to star, and  the formation of disks. For example, magnetic fields aligned with the rotation axis of the core can redistribute angular momentum through magnetic braking and subsequently inhibit the growth of protostellar disks \citep[][and references therein]{tsukamoto_review}. Nevertheless, the full role of magnetic fields on core scales remains an open question. Observing the change in the magnetic field orientation in the transition from cloud to core scales, and analyzing alignments between the core-scale magnetic field and dense core properties, will help us assess the dynamical importance of magnetic fields in the evolution of dense cores into stellar systems. 

Unfortunately, accurately estimating the magnetic field strength is difficult on core scales. We can more easily, however, measure the magnetic field orientation in the plane of the sky using the dust continuum emission. Elongated dust grains are preferentially oriented perpendicular to the local magnetic field, and therefore we can use the polarization emission from these dust grains to trace the local magnetic field \citep{Lazarian2007, 2007Lazarian_2, Andersson_2015}. 
Observations with the \textit{Planck} observatory are able to measure the polarization of dust emission at resolutions of 5$'$ (850 $\mu$m), revealing how the plane-of-sky magnetic field is oriented on angular scales comparable to large-scale structures in some star-forming clouds/regions (i.e., ``cloud scale''). On the other hand, observations with the James Clerk Maxwell Telescope (JCMT) probe dust polarization at a higher angular resolution of 14.$''$1 (850 $\mu$m), more comparable to the projected size of individual dense cores (i.e., ``core scale''). 

Molecular clouds tend to have ordered magnetic fields on cloud-scales, with a high degree of observed correlation on $\sim 10$~pc physical scales \citep{2016Planck, 2016Fissel, 2018Tahani}. Results from core scales, however, and the relationship between the cloud- and core-scale fields have been less conclusive. Comparisons between observations tracing cloud- and core-scale magnetic field orientations in star-forming regions such as NGC 1333 \citep{JCMTBISTRO_NGC1333_2020} and L1689 \citep{BISTRO_L1689} show that the fields across these scales are aligned with each other in some cores, while are perpendicular in others. 

Ideal magnetohydrodynamic (MHD) simulations by \cite{Chen2020} found that the magnetic field measured on core scales is generally aligned with the cloud-scale magnetic field, though there was a notable fraction of cores (approximately 10\%) with near perpendicular alignment. Simulated results on the alignment of other properties of the core and the core-scale magnetic field also vary. Some simulations show that the core-scale magnetic field is preferentially perpendicular to the core major axis and parallel to the core minor axis, and that there is no preferred alignment between the magnetic field and the angular momentum axes within cores \citep{chen_ostriker_2018}. \cite{Chen2020} found that in simulations, the major axes of cores in 3D space have a weak preference to be aligned perpendicular to the core-scale magnetic field, but have a random alignment overall to the large-scale magnetic field; cores identified in observations had significant differences in alignment to the large-scale magnetic field between regions. 
\cite{kuznetsova2020} also simulate the formation of dense cores during gravitational collapse in ideal MHD and, in contrast to the results discussed above, they find random alignment between the angular momentum axes of cores and the core-scale magnetic field. However, initial conditions vary between each of the simulations mentioned above and could impact the resulting relative orientations.
Overall, how the magnetic field varies in the transition from cloud to core scales and the alignment between core properties and the magnetic field remains unclear in observations, while simulations have a broad range of predictions depending on their initial conditions.

\cite{Pandhi2023} conducted a thorough analysis of 399 dense cores in nearby molecular clouds, identified in continuum emission from \textit{Herschel} and JCMT, and molecular line emission from the Green Bank Ammonia Survey (GAS). They found that there was no globally preferred orientation between the core major axis, core velocity gradient, and cloud-scale magnetic field, as measured by \textit{Planck} 5$'$ resolution polarized dust maps. For specifically proto-stellar cores, however, the elongation axis of the cores was uniquely preferentially perpendicular with the cloud-scale magnetic field\revision{. This is reminiscent of a picture where a protostellar disk is oriented perpendicular to the local magnetic field direction, and protostellar outflow extends along the local magnetic field direction. \cite{Pandhi2023} found no preferred orientation between the velocity gradient and the cloud-scale magnetic field however, as would have been expected if the elongated core was tracing a rotating protostellar disk. At different stages in the core collapse and disk formation, the velocity gradient would be some combination of infall and rotation motions, which may not be captured as a simple single linear velocity gradient. Regardless, the observed preferred relative orientation between the core elongation axis and the cloud-scale magnetic field suggests} that either: (i) the relative alignment between the core major axis and the cloud-scale magnetic field evolves from randomly aligned to perpendicular as starless and pre-stellar cores evolve into proto-stellar cores, or that (ii) this preferential anti-alignment might be the result of a selection effect where these cores are more likely to eventually evolve into protostellar cores \citep{Pandhi2023}. 

Higher resolution polarization data are needed to fully understand the role of magnetic fields in the evolution of dense cores. In this paper, we use observations of polarized dust emission from the JCMT B-Fields In STar-forming Region Observations (BISTRO) survey \citep{2017_WardThompson} and from legacy polarization catalogs \citep{2009SCUPOL} to measure the magnetic field at higher resolution (14\arcsec.1 vs. 5\arcmin\ for \textit{Planck}). Using these data, we analyze the orientation of the magnetic field in the transition from cloud to core scales, and we re-assess the correlations found by \cite{Pandhi2023} on cloud-scales, but now using the core-scale magnetic field.

The rest of the paper is organized as follows. In Section \ref{sec:data}, we describe the survey data, data reduction techniques, and resulting data products used in this work. In Section \ref{sec:results}, we present our results comparing the core-scale magnetic field orientation (JCMT) to the cloud-scale field (\textit{Planck}), and we also show our results comparing the orientation of core properties to core-scale magnetic fields. In Section \ref{sec:discussion}, we discuss our results in the context of previous observations and simulations. Finally, in Section \ref{sec:conclusion}, we provide a summary of our conclusions.

\section{Data and Reduction}\label{sec:data}

\subsection{Data} \label{subsec:data}

We use dust continuum maps from the JCMT via its polarimeter POL-2 \citep{POL-2_Bastien2011, 2007Lazarian_2, POL-2_Friberg2016}. We use data from the BISTRO survey for nearby star-forming regions, including Perseus B1 \citep{JCMTBISTRO_B1_2019}, NGC 1333 \citep{JCMTBISTRO_NGC1333_2020}, OMC-1 \citep{JCMTBISTRO_OMC1_2021}, and several regions in the Ophiuchus cloud complex \citep{BISTRO_OphA, BISTRO_OphB, BISTRO_OphC, BISTRO_L1689}. The survey data products have slightly different pixel scales between regions, but this does not affect our results going forward. The effective angular resolution of the JCMT at 850 micron is $\sim 14$\arcsec.1, a factor of over 20 improvement in resolution compared with similar \textit{Planck} observations. In the aforementioned regions, we obtain Stokes $I$, $Q$, and $U$ parameter maps, which we use to calculate polarization fraction and polarization angle maps. 

For regions outside of our BISTRO dataset for which we still have measured core properties, we additionally include in our analysis a catalog of polarization vectors \citep{2009SCUPOL} from the legacy Submillimetre Common-User Bolometer Array polarimeter \citep[SCUPOL;][]{SCUPOL_Greaves}, the previous polarization detector on the JCMT. These regions include IC348, L1448, L1455, OMC2/OMC3, NGC2024, and NGC2068 (at distances of 300--400pc). The SCUPOL maps have limited spatial coverage compared to the BISTRO maps, with polarization vectors typically extending about 1--2 arcminutes around the brightest continuum sources. Table \ref{tab:region_stats} lists the 14 regions in this study and their corresponding JCMT dataset. 

We also use 850$\mu$m (353 GHz) dust polarization observations from the \textit{Planck} collaboration to infer the cloud-scale magnetic field orientation \citep{2015Planck} for the clouds surrounding the aforementioned JCMT data (e.g., Perseus, Orion, Ophiuchus). As described in \cite{Pandhi2023}, the \textit{Planck} maps were smoothed to a resolution of 5\arcmin\ FWHM following the procedure in \cite{Soler2019}. 

The regions discussed in this paper range in distance from 138 parsecs (L1688) to 400 parsecs (Orion A) \citep{2018Ortiz-Leon, 2020Zucker}. For the JCMT data, this means an approximate physical resolution of about 0.01--0.03 parsecs, and for the \textit{Planck} data, an approximate physical resolution of about 0.2--0.7 parsecs. The average core \revision{radius} in these regions is $\sim$0.01--0.1 parsecs, hence the JCMT data measures magnetic field vectors on scales similar to or smaller than the typical core sizes. Therefore, for the rest of the paper, we use the terms ``core-scale magnetic field'' in reference to the magnetic field as derived from JCMT data, and ``cloud-scale magnetic field'' in reference to the magnetic field derived from \textit{Planck} data. Note that these definitions are independent from whether or not the vectors are physically coincident with a core. 

Finally, we obtain core properties, such as the velocity gradient orientation and the cloud-scale magnetic field orientation at the core, from the core catalog by \citet[][and references therein]{Pandhi2023}.\footnote{\url{https://www.canfar.net/citation/landing?doi=23.0008}} \revision{Core velocity gradients were taken from \cite{Pandhi2023} and were calculated from NH$_3$ emission from} the Green Bank Ammonia Survey \citep[GAS;][]{GAS_Friesen, Pineda_2026_GAS_DR2}. \revision{\cite{Pandhi2023} applied a 3-$\sigma$ threshold in the derived velocity gradient to ensure a robust measurement in gradient magnitude and orientation.} We obtain the \revision{deconvolved} major and minor axes and position angles of cores from core catalogs from the \textit{Herschel} Gould Belt Survey \citep[HGBS, 70--500 $\mu$m with a 36$''$ beam size;][]{HGBS_Andre} for the Perseus, Ophiuchus, and Orion B regions \citep{HGBSPerseus, HGBSOphiuchus, HGBSOrionB}. As there is currently no published core catalog from HGBS for Orion A, we use a core catalog of the Orion A region \citep{2025Pattle_JCMT} from the JCMT Gould Belt Survey \citep[JCMT GBS at 850 $\mu$m with a 14.1$''$ beam size;][]{JCMTGBS_WardThompson, JCMTGBS_Dempsey}. For both \textit{Herschel} and JCMT datasets, core catalogs were made using the \revision{\texttt{getsources}} and \revision{\texttt{getsf}} packages \citep{getsources, getsf}\revision{, which filters out larger-scale emission to produce a robust measurement of core properties, including major and minor axes. There were no uncertainties given on core position angles, so to ensure that the position angle is robust, \cite{Pandhi2023} applied a threshold of minor axis/major axis $\leq$ 0.9 to exclude cores that were not sufficiently elongated.} \revision{We take the cloud-scale magnetic field orientation at each core location from the catalog by \cite{Pandhi2023}. This was determined through dust polarization observations from the \textit{Planck} collaboration \citep{2015Planck}. For each core, the cloud-scale magnetic field orientation is given by the nearest \textit{Planck} vector to the center of the core, which is always within one half of the \textit{Planck} resolution, or 3$^\prime$ for this dataset.} 

\subsection{Methodology}\label{sec:methods}

With the Stokes $I$, $Q$, and $U$ maps and their respective uncertainty maps $\Delta I$, $\Delta Q$, and $\Delta U$ obtained from the BISTRO data, we calculate both the polarization fraction and the polarization angle as follows. Stokes $I$ is the total intensity, while $Q$ and $U$ capture the linear polarization in the 0--90$^{\circ}$ and 45--135$^{\circ}$ bases, respectively \citep{Pattle2019}. We then calculate the debiased polarized intensity $PI_\mathrm{deb}$ following the procedure described by \cite{JCMTBISTRO_B1_2019} and \cite{JCMTBISTRO_NGC1333_2020}:

\begin{equation}\label{eqn:PI}
    PI = \sqrt{Q^2 + U^2}
\end{equation}
\begin{equation}\label{eqn:PI_unc}
    \Delta PI = \frac{\sqrt{(Q^2 \cdot \Delta Q^2 + U^2 \cdot \Delta U^2)}}{PI}
\end{equation}
\begin{equation}\label{eqn:PI_debiased}
    PI_\mathrm{deb} = \sqrt{PI^2 - \Delta PI^2}
\end{equation}

With the debiased polarization intensity, we can then calculate maps for the polarization fraction $p$, the polarization angle $\theta$, and their respective uncertainties $\Delta p$ and $\Delta \theta$ as follows:

\begin{equation}\label{eqn:p}
    p = \frac{PI_\mathrm{deb}}{I}
\end{equation}
\begin{equation}\label{eqn:p_unc}
    \Delta p = p \cdot \sqrt{\left(\frac{\Delta PI}{PI_\mathrm{deb}}\right)^2 + \left(\frac{\Delta I}{I}\right)^2}
\end{equation}
\begin{equation}\label{eqn:theta}
    \theta = \frac{1}{2}\arctan{\frac{U}{Q}}
\end{equation}
\begin{equation}\label{eqn:theta_unc}
    \Delta \theta = \frac{1}{2} \frac{\sqrt{(Q \cdot \Delta U)^2 + (U \cdot \Delta Q)^2}}{PI_\mathrm{deb}^2}
\end{equation} 

After calculating these polarization parameters, we filter the BISTRO data with the following thresholds: $I / \Delta I > 5$ and $p / \Delta p > 3$. \revision{As a result, the typical uncertainty on the orientation of a single polarization vector is approximately 5$^\circ$.} We proceed by producing polarization vector catalogues of the remaining data, with each vector's location (RA, Dec), Stokes $I$, $Q$, and $U$ parameters, the polarization fraction $p$, the polarization angle $\theta$, and their corresponding uncertainties. As we cannot determine the direction of the polarization, we measure the polarization angles over a range of $-$90--90$^{\circ}$ (e.g., 45$^{\circ}$ and 225$^{\circ}$ are the same orientation). We run the same analysis as above on the \textit{Planck} data to infer the cloud-scale magnetic field.

We use the data directly from the \revision{legacy SCUPOL vector} catalog which was filtered as $I > 0$, $p / \Delta p > 2$, and $\Delta p < 4\%$. \revision{Given these thresholds, the typical reported uncertainty in the polarization vector orientation ranges between 4$^\circ-13^\circ$, slightly greater than that for BISTRO.} We find that the number of dense cores matched remains the same when using a higher signal-to-noise threshold of $p / \Delta p > 3$, and the mean magnetic field orientation and standard deviation of the magnetic field vectors in each region do not change significantly. We use these lower signal-to-noise thresholds on the data to retain more vectors, at the cost of slightly higher uncertainty in the polarization orientation. 

We \revision{match polarization vectors from the aforementioned catalogs to dense cores based on their properties} from core catalogs from HGBS, JCMT GBS, and \cite{Pandhi2023}. \revision{For each core, we create a core boundary ellipse given the core location, major and minor axes, and orientation. We then identify polarization vectors that fall within the ellipse.} Most cores that fall within the coverage of the JCMT data are matched with at least one vector, though the number of vectors matched to each core varies significantly, ranging from just 2 vectors to over 100. Cores that do not have any vector matches were either outside of the range of the BISTRO polarization data, or were in regions that did not meet our polarization fraction sensitivity thresholds. We successfully matched a total of 59 cores with the BISTRO data across NGC 1333 (16 cores), B1 (6 cores), OMC-1 (12 cores), and various regions in the Ophiuchus cloud complex (25 cores). Further, we matched a further 20 cores within various star-forming regions in the Perseus, Orion A, and Orion B clouds to the polarization vectors from the SCUPOL data. Our total sample consists of 79 cores with matched polarization vector data, with a range in radii from 0.006 to 0.054 pc and a median core radius of 0.022 pc. 

We convert the angles from the polarization orientation to the magnetic field orientation, noting that the polarization is perpendicular to the magnetic field \citep{Lazarian2007}. We also convert our angles to a range from 0 to 180 degrees (in east from north) to match the convention used by \cite{Pandhi2023}. \revision{For each core, we calculate the mean magnetic field orientation by taking a circular average over all the magnetic field vectors matched to the core. This factors} in wrap-around effects\revision{, ensuring that} the average of two angles at 2 and 178 degrees \revision{is} 0 degrees, rather than 90 degrees.

\begin{figure*}
    \centering
    \includegraphics[width=\textwidth, trim={0cm 0cm 0cm 0cm}, clip]{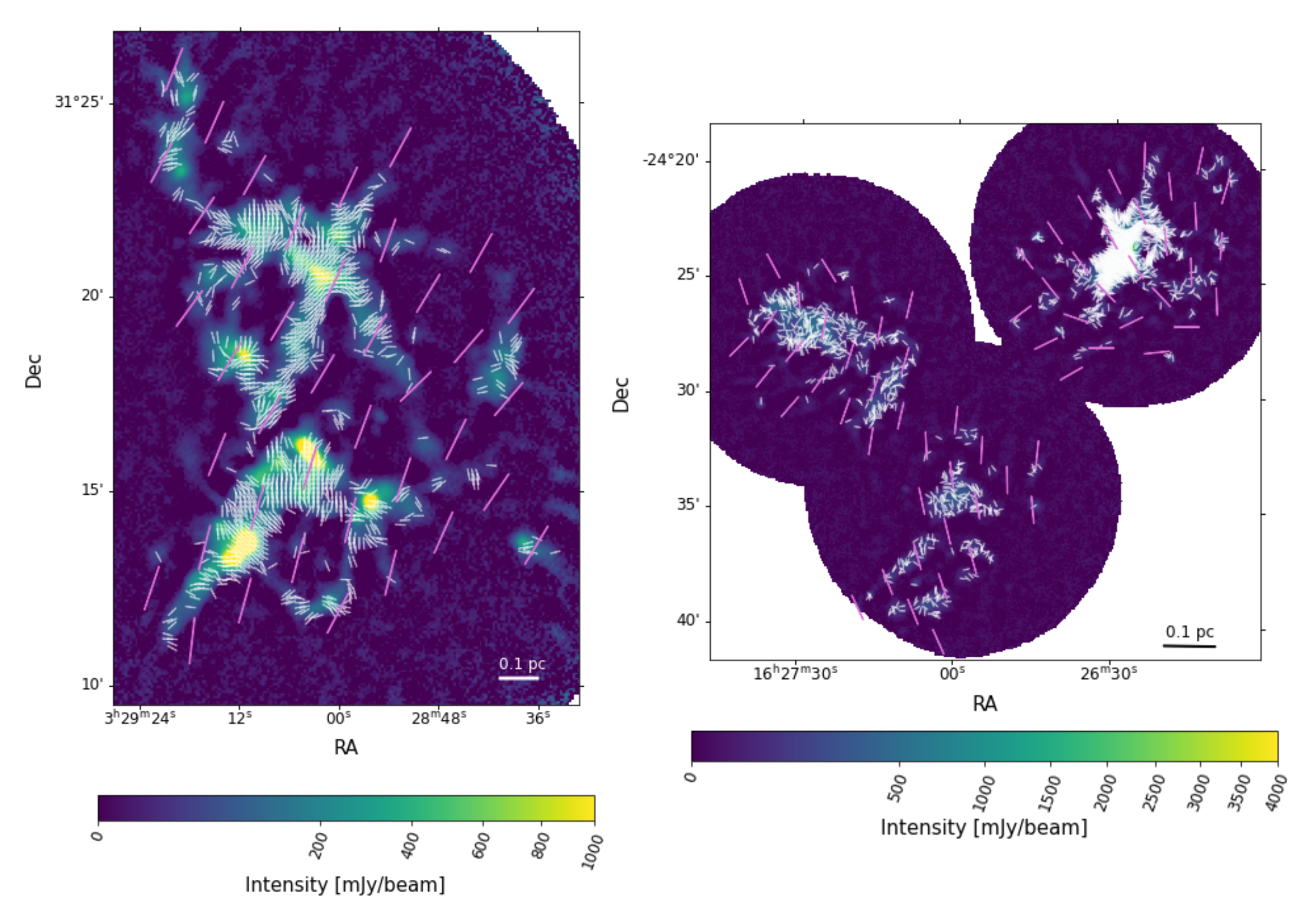}
    \caption{Left: NGC 1333 in 850 $\mu$m continuum emission from JCMT (colourscale) overlaid with the magnetic field orientation at FWHM 14.1$''$ \citep[white vectors;][]{JCMTBISTRO_NGC1333_2020} and the magnetic field orientation at FWHM 5$'$ \citep[orchid vectors;][]{2016Planck}. Right: L1688 (Oph A, Oph B, Oph C/E/F) presented using the same setup. Maps of the other regions covered in this paper are found in Appendix \ref{appendix:vectors}.
    \label{fig:ngc1333_visual_results}}
\end{figure*}

Finally, we convert the cloud-scale magnetic field vectors from Galactic coordinates to equatorial coordinates, in order to directly compare to our core-scale results. For each vector, we do this by drawing a unit vector pointing to Galactic north from the origin of the vector, converting the endpoints to equatorial coordinates, and then calculating the angle between the converted Galactic north vector and a unit vector pointing to equatorial north drawn from the same location in equatorial coordinates. We then add or subtract this angle from the vector orientation in Galactic coordinates as required to adjust to equatorial coordinates.

Our data processing results in a final core catalog that includes each core's sky position, major/minor axis, core orientation (\revision{or core position angle,} $\theta_{C}$), velocity gradient orientation (\revision{direction of the velocity gradient, }$\theta_{G}$), core-scale magnetic field orientation in equatorial coordinates as found through JCMT data ($\theta_{B}$), and the cloud-scale core magnetic field orientation also in equatorial coordinates from \textit{Planck} from the \cite{Pandhi2023} catalog ($\theta_\mathrm{Planck}$). In total, our core catalog spans 14 different star-forming regions in 3 cloud complexes.

\begin{deluxetable*}{lcccccccc} 
\tabletypesize{\footnotesize} 
\tablecolumns{9} 
\tablecaption{Table displaying the mean and standard deviation of all magnetic field vector orientations across each star-forming region, from both JCMT and \textit{Planck}, organized by cloud. Angles are shown in degrees, and are measured east of north in equatorial coordinates.   \label{tab:region_stats}} 
\tablehead{ 
\colhead{Cloud} & \colhead{Region} & \colhead{Dataset \tablenotemark{a}} & \colhead{$\theta_\mathrm{JCMT}$} & \colhead{$\sigma_\mathrm{JCMT}$} & \colhead{$\#_\mathrm{JCMT}$ \tablenotemark{b}} & \colhead{$\theta_\mathrm{Planck}$} & \colhead{$\sigma_\mathrm{Planck}$} & \colhead{$\#_\mathrm{Planck}$ \tablenotemark{b}} \\ 
\colhead{} & \colhead{} & \colhead{} & \colhead{($^{\circ}$)} & \colhead{($^{\circ}$)} & \colhead{} & \colhead{($^{\circ}$)} & \colhead{($^{\circ}$)} & \colhead{}
} 
\startdata 
Perseus & NGC1333 & BISTRO & 134 & 58 & 1280 & 151 & 9 & 40\\ 
 & B1 & BISTRO & 167 & 20 & 324 & 160 & 9 & 22 \\ 
 & IC348 & SCUPOL & 172 & 19 & 49 & 162 & 2 & 2 \\ 
 & L1448 & SCUPOL & 167 & 57 & 130 & 166 & 5 & 8 \\ 
 & L1455 & SCUPOL & 74 & 27 & 59 & 159 & 8 & 5 \\  
Ophiuchus & Oph A & BISTRO & 68 & 38 & 1319 & 26 & 43 & 28 \\
 & Oph B & BISTRO & 26 & 79 & 408 & 165 & 24 & 20 \\
 & Oph C/E/F & BISTRO & 38 & 63 & 241 & 13 & 11 & 17 \\
 & L1689-1 & BISTRO & 38 & 36 & 540 & 17 & 4 & 12 \\
 & L1689-2 & BISTRO & 84 & 46 & 168 & 11 & 9 & 10 \\
Orion A & OMC1 & BISTRO & 115 & 42 & 7551 & 123 & 12 & 46 \\ 
 & OMC2/OMC3 & SCUPOL & 45 & 43 & 361 & 81 & 22 & 30 \\ 
Orion B & NGC2024 & SCUPOL & 63 & 42 & 203 & 169 & 54 & 8 \\ 
 & NGC2068 & SCUPOL & 11 & 40 & 285 & 5 & 11 & 13 \\
\enddata 
\tablenotetext{a}{Denotes whether JCMT data is from BISTRO or SCUPOL}
\tablenotetext{b}{Total number of magnetic field vectors in the JCMT/Planck data after applying selection criteria}
\end{deluxetable*}

\begin{figure*}
    \centering
    \includegraphics[width=\textwidth, trim={0cm 0cm 0cm 0cm}, clip]{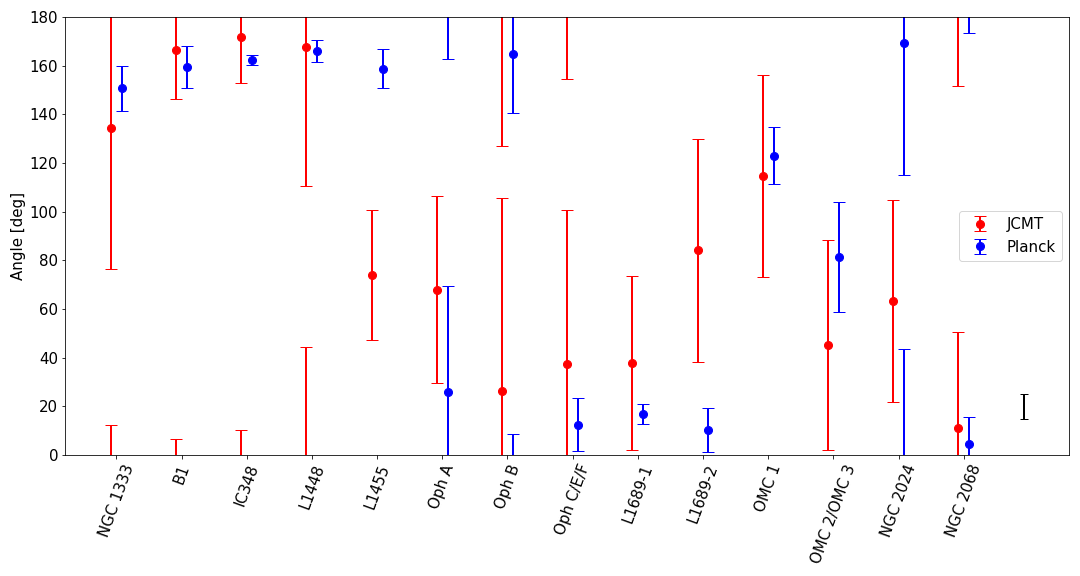}
    \caption{Mean magnetic field orientation and standard deviation for each star-forming region. JCMT data are shown in red, while \textit{Planck} data are shown in blue. The mean of both distributions is shown with a solid point, with the standard deviation represented by the span of its respective error bar. The black error bar on the lower right hand side represents the average uncertainty ($\sim 5^\circ$) of an individual core-scale vector in the JCMT data. Full information can be found in Table \ref{tab:region_stats}.}
    \label{fig:error_bar_plot}
\end{figure*}

\newpage

\section{Results}\label{sec:results}

We first discuss our results comparing the magnetic field orientation measured with the JCMT to the cloud-scale field observed by \textit{Planck} (Section \ref{subsec:results_field}). We then discuss our results comparing the core-scale magnetic field orientations to the properties of cores in our matched core catalog (Section \ref{subsec:results_cores}). 

\subsection{Transition from cloud-scale to core-scale magnetic field} \label{subsec:results_field}

We first examine the magnetic field orientation observed with the JCMT and \textit{Planck} across each of the star-forming regions studied in this work. Figure \ref{fig:ngc1333_visual_results} features an example of the magnetic field vectors for NGC 1333 and for L1688, showing the JCMT 850$\mu$m continuum emission maps of NGC 1333 and L1688 with the JCMT/BISTRO and \textit{Planck} magnetic field vectors overlaid. We show \textit{Planck} vectors only in areas overlapping the BISTRO vectors. We refer the reader to Appendix \ref{appendix:vectors} for similar maps of the other regions. Qualitatively, we see a complex core-scale magnetic field in comparison to the largely smooth and uniform cloud-scale magnetic field.

Table \ref{tab:region_stats} gives the circular average magnetic field angle and its standard deviation from the JCMT data (BISTRO or SCUPOL) and \textit{Planck} data for each of the 14 regions in this study. We also include the number of magnetic vectors used in the circular average. All angles are given in degrees and are measured east of north in equatorial coordinates. Note that the standard deviation is not a perfect metric of the spread around the mean, as some regions have non-Gaussian distributions. Nevertheless, the standard deviations of the core-scale magnetic field are greater than the average uncertainty of an individual core-scale vector (typically $\sim5^\circ$ based on Equation \ref{eqn:theta_unc}), indicating that the standard deviations are a real representation of the distribution of angles and not due to measurement uncertainties. Figure \ref{fig:error_bar_plot} shows the mean core-scale and cloud-scale orientations for each region with error bars representing the standard deviation.

\begin{figure*}[!ht]
    \centering
    \includegraphics[width=\textwidth, trim={1cm 41cm 1cm 2cm}, clip]{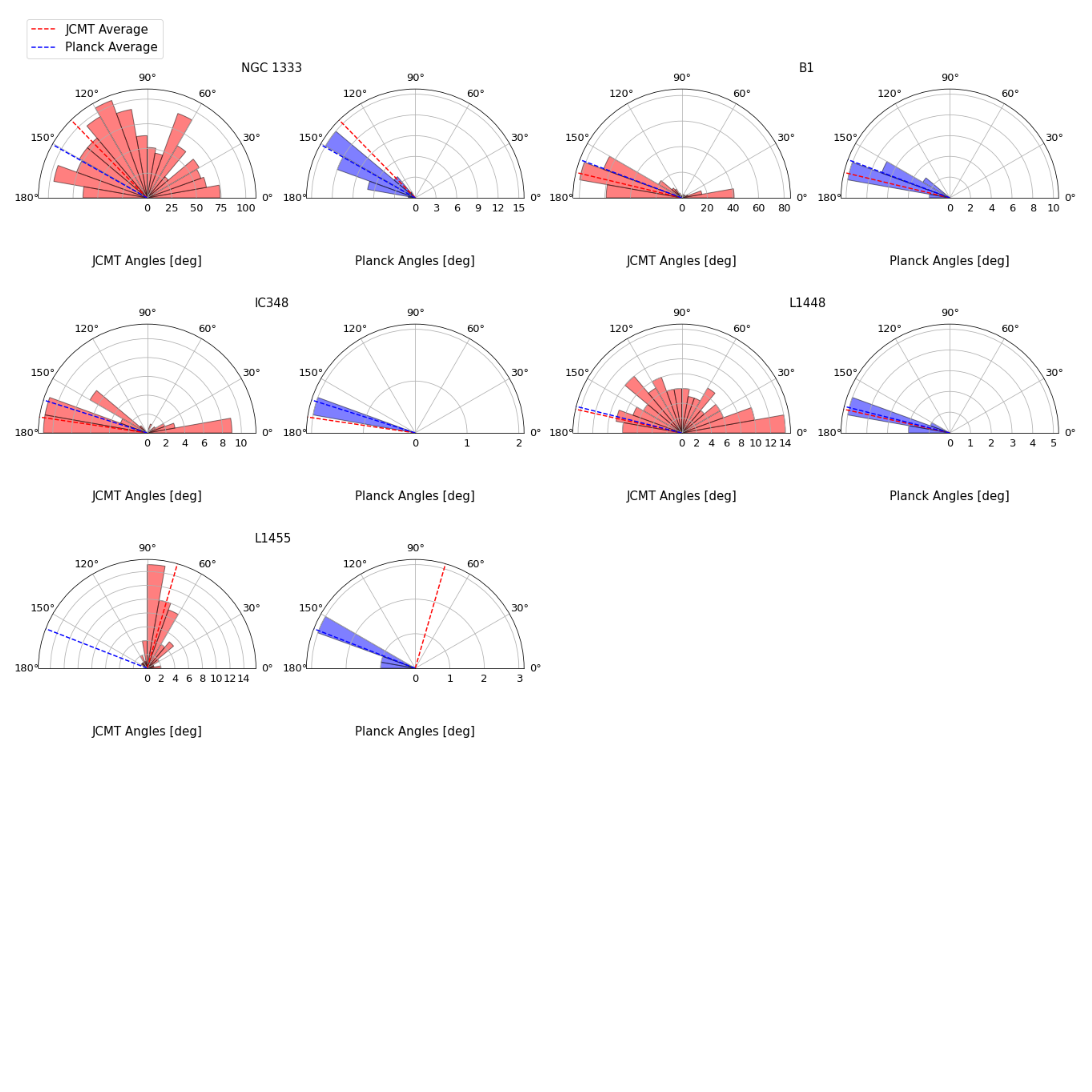}
    \caption{Polar histograms of all magnetic field vector orientations for each star-forming region in the Perseus molecular cloud. JCMT data are shown in red, while \textit{Planck} data are shown in blue. The mean of both distributions is shown with a red and blue dotted line, respectively, on both histograms. Full information can be found in Table \ref{tab:region_stats}.}
    \label{fig:perseus_polar_results}
\end{figure*}

\begin{figure*}
    \centering
    \includegraphics[width=\textwidth, trim={1cm 41cm 1cm 2cm}, clip]{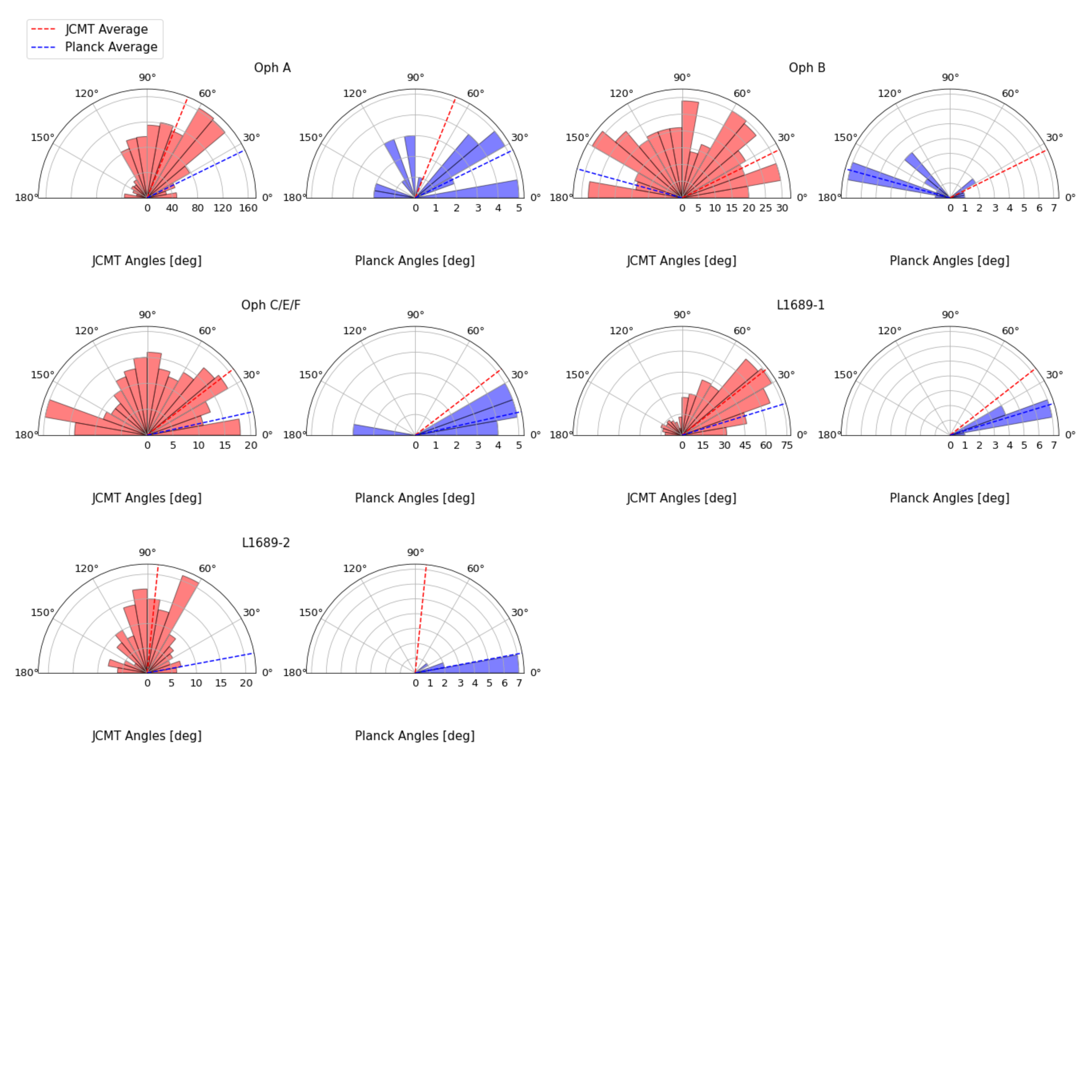}
    \caption{Same as Figure \ref{fig:perseus_polar_results} but for Ophiuchus.}
    \label{fig:oph_polar_results}
\end{figure*}

\begin{figure*}
    \centering
    \includegraphics[width=\textwidth, trim={1cm 70cm 1cm 2cm}, clip]{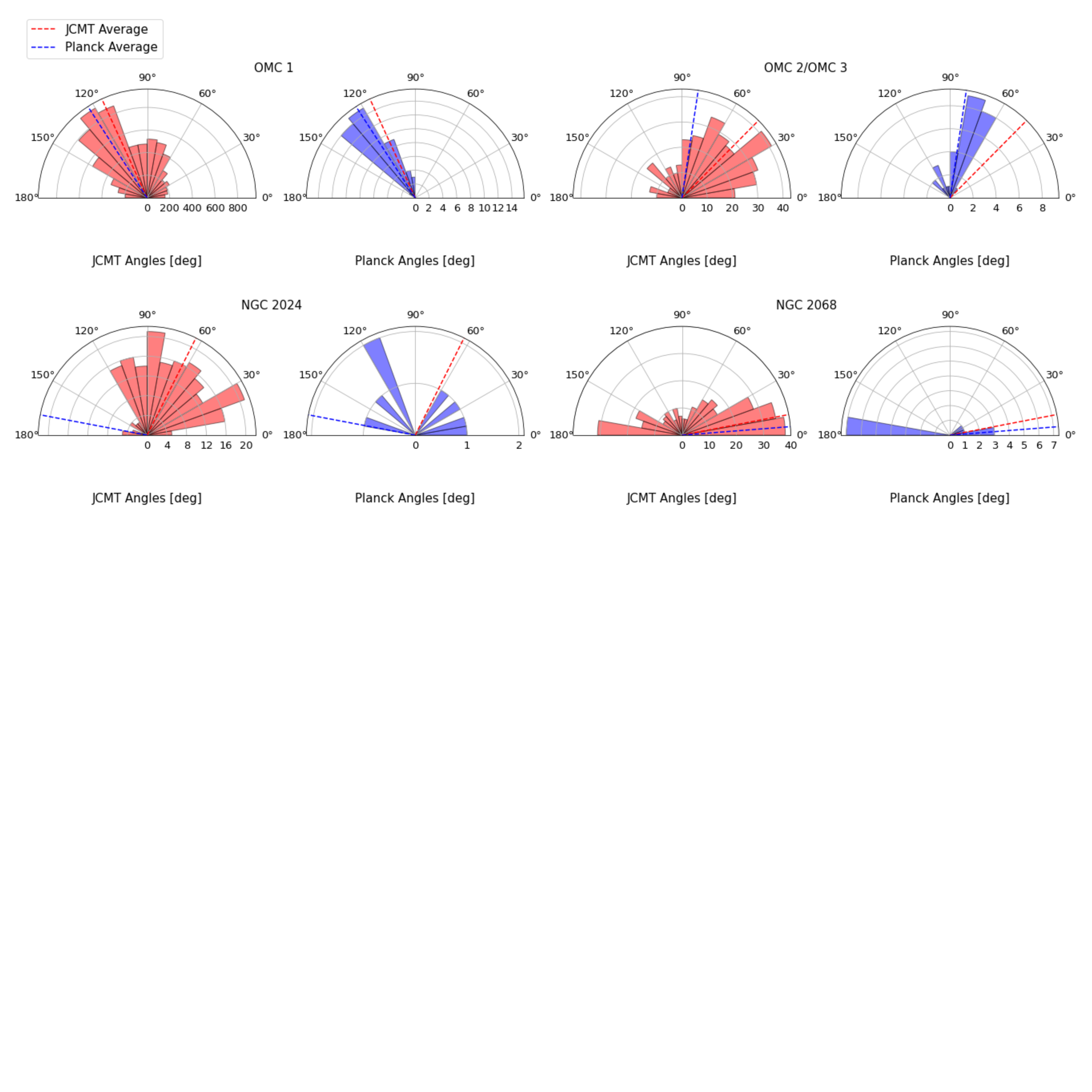}
    \caption{Same as Figure \ref{fig:perseus_polar_results} but for Orion A and Orion B.}
    \label{fig:orion_polar_results}
\end{figure*}

To better compare the alignment between the core-scale and cloud-scale magnetic fields for each region, we plot polar histograms using the full set of JCMT and \textit{Planck} magnetic field vectors in Figures \ref{fig:perseus_polar_results}--\ref{fig:orion_polar_results}. As with Figure \ref{fig:ngc1333_visual_results}, we only consider \textit{Planck} vectors that overlap with the JCMT data.

Overall, there is much greater spread in the orientation of the core-scale magnetic field than in the cloud-scale field. We see this greater spread in the standard deviation (Figure \ref{fig:error_bar_plot} and Table \ref{tab:region_stats}) and in the distribution of angles in the polar plots (Figures \ref{fig:perseus_polar_results}--\ref{fig:orion_polar_results}). The only region where the core-scale field has less spread is NGC 2024, where the \textit{Planck} data show an equivalent spread in angle to the JCMT data (see Figure \ref{fig:orion_polar_results}), although the statistics are limited in this region with only 8 \textit{Planck} vectors. These differences are not unsurprising when looking at the magnetic field vectors in the individual maps. For example, Figure \ref{fig:ngc1333_visual_results} shows that the core-scale magnetic field in NGC 1333 is much more complex in comparison to the smooth and uniform cloud-scale magnetic field (see Appendix \ref{appendix:vectors} for the other regions) even though the average field orientation matches between core-scale and cloud-scale vectors (see Figure \ref{fig:error_bar_plot} and Table \ref{tab:region_stats}). This contrast between the core-scale and cloud scale magnetic field is not seen in all the regions. In particular, B1 and L1448 generally show good agreement between their core- and cloud-scale magnetic fields (see Figure \ref{fig:perseus_visual} in Appendix \ref{appendix:vectors}).

We also see a large variety in alignment between the core-scale and cloud-scale magnetic fields between the regions. B1, IC348, and L1455 have distinctively smaller spreads in the core-scale magnetic field among the regions we studied (see Figure \ref{fig:perseus_polar_results}), with standard deviations less than 30$^\circ$. We also see that the average orientation among the core-scale and cloud-scale magnetic fields has varying degrees of alignment between regions, ranging from nearly perfectly aligned in regions such as L1448, to almost perpendicular in regions such as L1455. We also note that the regions where the core-scale magnetic field vectors have a small spread do not necessarily align with the regions where the core-scale and cloud-scale magnetic field align well. For example, while both B1 and L1455 display a small spread among the core-scale magnetic field vectors, B1 is fairly well aligned between the core-scale and cloud-scale field, while L1455 displays an almost perpendicular alignment.

We can sort the regions into three distinct cases of alignment between the core-scale and cloud-scale magnetic fields.

\begin{itemize}
  \item \textbf{Case 1: Aligned and Ordered.} These are regions in which the mean core-scale and cloud-scale magnetic field orientations align within $30^\circ$, and the standard deviation of the two distributions fall under $30^\circ$. \revision{We find only two of 14 total regions classified as Case 1: B1 and IC348, both in the Perseus molecular cloud.} 
  \item \textbf{Case 2: Aligned and Disordered.} These are regions in which the mean core-scale and cloud-scale magnetic field orientations align within $30^\circ$, but the standard deviation in the JCMT data is greater than $30^\circ$. \revision{Six of the 14 total regions are classified in Case 2, spanning several molecular clouds: NGC 1333 and L1448 in Perseus; Oph C/E/F and L1689-1 in Ophiuchus; and OMC 1 and NGC 2068 in Orion A and B, respectively. }
  \item \textbf{Case 3: Misaligned.} These are regions where the mean core-scale and cloud-scale magnetic field orientations do not align within $30^\circ$. 
  \revision{The remaining six regions, L1455, Oph A, Oph B, L1689-2, OMC2/OMC3, and NGC 2024, are classified as Case 3.
  Toward all but one of these regions, the standard deviation in the core-scale magnetic field is higher than 30$^\circ$, indicating a disordered field. The remaining region, L1455, only falls just short of this criteria, however, with a standard deviation of 27$^\circ$. }
\end{itemize}

These classifications do not appear to be impacted by whether the core-scale magnetic field orientations are derived from BISTRO or SCUPOL data.

\begin{figure*}
    \centering
    \includegraphics[width=\textwidth, trim={0cm 0cm 0cm 0cm}, clip]{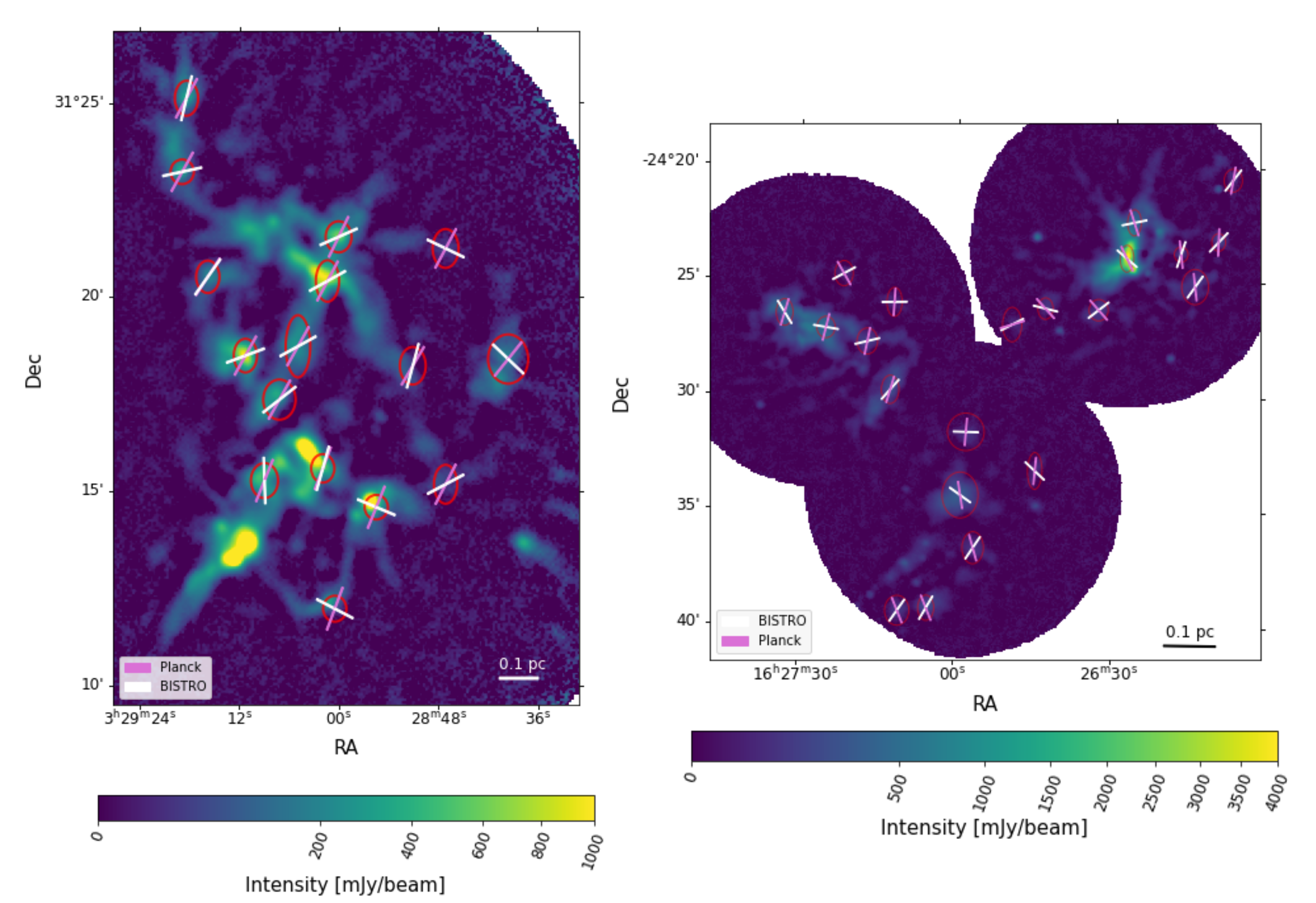}
    \caption{Left: NGC 1333 with colourscale as in Figure \ref{fig:ngc1333_visual_results}, with the mean magnetic field orientation of cores as derived from BISTRO (in white) and from \textit{Planck} \citep[in orchid;][]{Pandhi2023}. Core shapes are shown with ellipses (in red). Right: L1688 (Oph A, Oph B, Oph C/E/F) presented using the same setup. Maps of the other regions covered in this paper can be found in Appendix \ref{appendix:cores}.}
    \label{fig:ngc1333_core_results}
\end{figure*}

\subsection{The magnetic field orientation of cores} \label{subsec:results_cores}

In this subsection, we \revision{investigate the relative orientation between $\theta_B$ (the core-scale magnetic field orientation), $\theta_\mathrm{Planck}$ (the cloud-scale magnetic field orientation), $\theta_C$ (the core orientation), and $\theta_G$ (the core velocity gradient orientation) of the cores}. Here, we assume that the majority of the polarized emission as measured by the JCMT is from the core, but we acknowledge that there may be a smaller contribution from the parent filament and cloud, since the polarized dust emission is optically thin at these wavelengths. As in the previous section, we plot in Figure \ref{fig:ngc1333_core_results} the continuum emission map of NGC 1333 and L1688, now overlaid instead with the matched cores found for each region (core major/minor axes and position angles provided by HGBS and JCMT GBS; see Section \ref{sec:data}). See Appendix \ref{appendix:cores} for all remaining regions. For each core, we show its overall core-scale and cloud-scale magnetic field orientation. 

We find a variety of alignments between the core-scale and cloud-scale magnetic field orientations within cores. Figure \ref{fig:ngc1333_core_results} shows some core- and cloud-scale magnetic field orientations are nearly perfectly aligned, others are almost perpendicular, and some are neither preferentially parallel or perpendicularly aligned. 

\begin{figure*}
    \centering
    \includegraphics[width=\textwidth, trim={4.5cm 0cm 4.5cm 1.5cm}, clip]{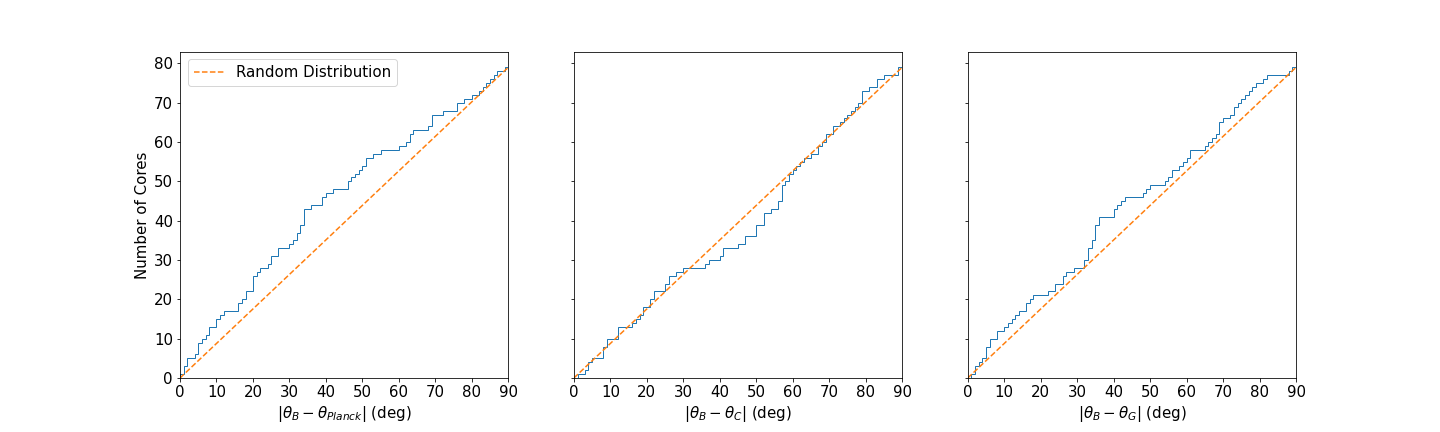}
    \caption{Cumulative distribution functions of $|\theta_{B} - \theta_\mathrm{Planck}|$ (the relative orientation between the JCMT (core-scale) and \textit{Planck} (cloud-scale) magnetic field orientation at each core), $|\theta_{B} - \theta_C|$ (the relative orientation between the core-scale magnetic field and the core orientation), and $|\theta_{B} - \theta_G|$ (the relative orientation between the core-scale magnetic field and the core velocity gradient orientation) for all cores. The dashed line indicates a completely random distribution of alignment between the two vectors.}
    \label{fig:global_cumulative_histograms}
\end{figure*}

We confirm the different distributions between the overall core- and cloud-scale magnetic field orientation within cores by using the two-sample Anderson-Darling \citep[AD;][]{ad_test} and Kolmogorov-Smirnov \citep[KS;][]{Kolmogorov1933, smirnov1948, ks_test} tests. The Anderson-Darling test statistic and p-value (which is floored/capped at 0.001/0.25) and the Kolmogorov-Smirnov test statistic and p-value are reported in Table \ref{tab:ad_ks_tests_1}. We use a p-value of $\leq$ 0.05 for each test to reject the null hypothesis that the two samples are drawn from the same underlying distribution at a 95\% confidence. The tests show that the core- and cloud-scale magnetic field data are unlikely to be drawn from the same underlying distribution. See the column labeled $\theta_{B}$ vs. $\theta_\mathrm{Planck}$ in Table \ref{tab:ad_ks_tests_1} for the results.

\begin{deluxetable}{lccc}
\tabletypesize{\footnotesize} 
\tablecolumns{4} 
\tablewidth{1.0\columnwidth} 
\tablecaption{Summary of Anderson-Darling (A-D) and Kolmogorov–Smirnov (K-S) tests on whether $\theta_B$ and $\theta_\mathrm{Planck}$ is consistent with being drawn from the same distribution. The results are shown for all cores in the first row, and in star-forming regions with more than 10 cores for the following rows. Each entry on the table represents the resultant p-value of the test. Results that pass the $\geq$ 95\% significance threshold (p $\leq$ 0.05) and therefore reject the null hypothesis are presented in boldface. \label{tab:ad_ks_tests_1}}
\tablehead{ 
\colhead{Region} & \colhead{\# of Cores} & \multicolumn{2}{c}{$\theta_{B}$ vs. $\theta_\mathrm{Planck}$} \\ 
\colhead{} & \colhead{} & \colhead{A-D Test} & \colhead{K-S Test}
} 
\startdata 
All Cores & 79 & \textbf{$\leq$ 0.001} & \textbf{0.00008} \\ 
NGC 1333 & 16 & \textbf{$\leq$ 0.001} & \textbf{0.0001} \\ 
L1688 & 21 & \textbf{0.003} & \textbf{0.02} \\
OMC 1 & 12 & \textbf{0.02} & \textbf{0.03} \\ 
OMC 2/OMC 3 & 11 & \textbf{0.005} & \textbf{0.004} \\ 
\enddata 
\end{deluxetable}

\begin{deluxetable*}{lccccccc} 
\tabletypesize{\footnotesize} 
\tablecolumns{8} 
\tablewidth{0pt} 
\tablecaption{Summary of Anderson-Darling (A-D) and Kolmogorov–Smirnov (K-S) tests on whether the relative orientation between $\theta_B$ and $\theta_\mathrm{Planck}$, $\theta_B$ and $\theta_C$, and $\theta_B$ and $\theta_G$ are consistent with being drawn from a random distribution between $0^{\circ} \leq \theta \leq 90^{\circ}$. The results are shown for all cores in the first row, and in star-forming regions with more than 10 cores for the following rows. Each entry on the table represents the resultant p-value of the test. Results that pass the $\geq$ 95\% significance threshold (p $\leq$ 0.05) and therefore reject the null hypothesis are presented in boldface.   \label{tab:ad_ks_tests_2}} 
\tablehead{ 
\colhead{Region} & \colhead{\# of Cores} & \multicolumn{2}{c}{$|\theta_B - \theta_\mathrm{Planck}|$} & \multicolumn{2}{c}{$|\theta_B - \theta_C|$} & \multicolumn{2}{c}{$|\theta_B - \theta_G|$} \\ 
\colhead{} & \colhead{} & \colhead{A-D Test} & \colhead{K-S Test} & \colhead{A-D Test} & \colhead{K-S Test} & \colhead{A-D Test} & \colhead{K-S Test}
} 
\startdata 
All Cores & 79 & $\geq$ 0.25 & 0.32 & $\geq$ 0.25 & 0.82 & $\geq$ 0.25 & 0.82 \\ 
NGC 1333 & 16 & $\geq$ 0.25 & 0.72 & $\geq$ 0.25 & 0.95 & $\geq$ 0.25 & 0.95 \\ 
L1688 & 21 & $\geq$ 0.25 & 0.60 & $\geq$ 0.25 & 0.99 & $\geq$ 0.25 & 0.85 \\
OMC 1 & 12 & $\geq$ 0.25 & 0.87 & 0.17 & 0.26 & $\geq$ 0.25 & 0.998 \\ 
OMC 2/OMC 3 & 11 & $\geq$ 0.25 & 0.997 & $\geq$ 0.25 & 0.997 & 0.10 & 0.21 \\  
\enddata 
\end{deluxetable*}

Though we conclude that the two distributions are statistically distinct, we can test whether there is a preferred relative orientation between the core- and cloud-scale magnetic field orientations. In Figure \ref{fig:global_cumulative_histograms} (left panel), we plot a cumulative histogram of the differences between the JCMT and \textit{Planck} datasets $|\theta_{B} - \theta_\mathrm{Planck}|$. We test whether $|\theta_{B} - \theta_\mathrm{Planck}|$ for the observed cores are consistent with being drawn from a random distribution with the same tests as before. To do this, we compare $|\theta_{B} - \theta_\mathrm{Planck}|$ with a dataset of the same size with a uniform distribution in angles from 0--90$^{\circ}$ (as the samples represent the difference between two orientations, they will only fall between a range of 0--90$^{\circ}$). The expected cumulative histogram for the random distribution is shown as the dashed line in Figure \ref{fig:global_cumulative_histograms}. We use the same thresholds to conclude significance as the previous tests. In this case, our null hypothesis is that $|\theta_{B} - \theta_\mathrm{Planck}|$ is consistent with being drawn from a random distribution. See Table \ref{tab:ad_ks_tests_2} for the results. For the entire core sample, we find that $|\theta_{B} - \theta_\mathrm{Planck}|$ is consistent with being drawn from a random distribution. In a cumulative histogram like those shown in Figure \ref{fig:global_cumulative_histograms}, a distribution that is preferentially more parallel will show a shift toward the upper left, while a distribution that is preferentially more perpendicular will show a shift toward the lower right. For an illustrative example, see Figure 8 in \cite{2018Sadavoy}. While Figure \ref{fig:global_cumulative_histograms} (left) shows that $|\theta_{B} - \theta_\mathrm{Planck}|$ appears to fall above the expectation for a random distribution (as expected for a preferentially parallel relative orientation), the test results show that this is not significant and we cannot reject the null hypothesis that the distribution is consistent with random.

We run the same analysis on the alignment between the core-scale magnetic field and two key core properties: the core orientation $\theta_{C}$ (characterized by the major axis of the core) and the core velocity gradient orientation $\theta_{G}$. These two core properties are provided by the core catalog by \cite{Pandhi2023}. We plot cumulative histograms of the relative orientation between the core-scale magnetic field and the core orientation ($|\theta_{B} - \theta_{C}|$) in Figure \ref{fig:global_cumulative_histograms} (middle), and the relative orientation between the core-scale magnetic field and the core velocity gradient orientation ($|\theta_{B} - \theta_{G}|$) in Figure \ref{fig:global_cumulative_histograms} (right). \revision{Typically, the velocity gradient is often assumed to be due to rotation in the core. There would likely be other bulk motions contributing to the observed velocity gradient, however, measurements of core velocity gradients remain good measures of the angular momentum orientation \citep{2000Burkert}. In a magnetically-dominated scenario, you would expect the core to collapse along the field lines with the rotation direction aligned with the core elongation, and perpendicular to the magnetic field lines.}  

Table \ref{tab:ad_ks_tests_2} lists the results of the AD and KS tests for both. The test results show that both $|\theta_{B} - \theta_{C}|$ and $|\theta_{B} - \theta_{G}|$ are consistent with being drawn from a random distribution. As \cite{Pandhi2023} previously found a slight preference for protostellar cores to be aligned perpendicular with the cloud-scale magnetic field (for 113 cores out of 329), we run the same analysis with our sample of 31 protostellar cores (out of 79 cores) and the core-scale magnetic field. However, the test results for our smaller sample find that the alignment between the core-scale magnetic field and the core orientation of protostellar cores is consistent with being random. We find the same results for prestellar and starless cores as well. We discuss this further in Section \ref{sec:discussion}.

We split the full core catalog by star-forming region to run the same analysis on specific regions. We limit this analysis to regions with at least 10 cores to ensure that we have sufficient numbers of cores for statistical analysis. We choose to treat the closely-grouped Oph A, Oph B, and Oph C/E/F (previously analyzed separately) as one region for this analysis, as they do not have enough cores individually to meet the threshold, and together they form the L1688 star-forming region. This leaves us with four regions: NGC 1333, L1688, OMC1, OMC2/OMC3. Table \ref{tab:ad_ks_tests_2} presents the AD and KS test results for each of the regions. From the test results, we conclude that the core- and cloud-scale magnetic fields are unlikely to be drawn from the same underlying distribution for each of the regions. We also conclude that for each of the regions, $|\theta_{B} - \theta_\mathrm{Planck}|$, $|\theta_{B} - \theta_{C}|$, and $|\theta_{B} - \theta_{G}|$ are consistent with being drawn from random distributions.

The final core catalog produced for the 79 cores analyzed in this paper includes information about the average core-scale magnetic field orientation $\theta_{B}$, standard deviation of the core-scale magnetic field orientations of all vectors matched to the core, number of vectors matched to the core, and alignments between $\theta_{B}$ and $\theta_\mathrm{Planck}$, $\theta_{C}$, and $\theta_{G}$. The full table can be found at Table \ref{tab:core_catalogue} in Appendix \ref{appendix:core_catalogue}.

\section{Discussion}\label{sec:discussion}

\subsection{Projection effects} \label{subsec: projection}

\revision{The magnetic field vectors and core properties (and hence the relative orientation between the two) analyzed in this work are measured as 2D plane-of-sky projections of their underlying 3D orientations; this projection effect is important to discuss and quantify in the context of our results. In individual cores, it can be difficult to determine the 3D magnetic field orientation from 2D observations; e.g., \cite{2025Tritsis} demonstrated that hourglass-shaped morphologies in core magnetic fields cannot easily be discerned unless the core is viewed edge on, or close to it. On a population level, however, \cite{2000Basu} showed that a core with a magnetic field aligned parallel to its minor axis viewed from different angles has a distribution of relative orientations between the projected magnetic field and minor axis that is peaked at 0$^\circ$ and has an average offset of only 10--30$^\circ$. Supporting this point, \cite{Chen2020} used synthetic 2D observations of 3D simulations to show that the relative alignment between dense cores and the background magnetic field does not significantly depend on the viewing angle. Hence, measurements of the 2D relative orientation between the magnetic field and core properties should still correctly diagnose an underlying preferred orientation in 3D space between the two given a large sample size of cores, such as the one presented in this work.}

\subsection{Alignment between cloud-scale and core-scale magnetic field}\label{sec:discussion_field}

\revision{The \textit{Planck} data probes angular scales greater than its resolution of 5\arcmin, while observations with POL-2 at the JCMT are limited to scale sizes up to 80\arcsec{} due to the instrument's mapping speed of 8\arcsec/s (e.g., see \cite{2013Chapin}).\footnote{\url{https://www.eaobservatory.org/jcmt/instrumentation/continuum/scuba-2/pol-2/}}\footnote{\url{https://www.eaobservatory.org/jcmt/instrumentation/continuum/scuba-2/observing-modes/}}
Our results in Section \ref{sec:results} suggest that the cloud-scale magnetic field often appears inconsistent with being simply a smoothed version of the core-scale field.
To test quantitatively whether the \textit{Planck} data is consistent with the BISTRO data smoothed to larger angular scales, we smooth the BISTRO Stokes I, Q and U maps in the NGC 1333, B1, OMC 1, and L1688/L1689 regions to an angular resolution of 2\arcmin, and recalculate the magnetic field orientation following Section \ref{sec:methods}.} 

\revision{We find that smoothing does not lead to alignment in cases where the BISTRO vectors, at their original resolution, were already misaligned to the \textit{Planck} vectors, or in cases where they were disordered. In general, the magnetic field orientation derived from the smoothed BISTRO data in these regions remains misaligned to those from the \textit{Planck} data, with the exceptions of B1 (where the two datasets already showed alignment at the original resolution), and OMC 1 (where the mean fields were aligned at the original resolution but the BISTRO vectors were disordered). In all other regions tested, AD and KS tests show that the smoothed datasets are not drawn from the same underlying distribution as the \textit{Planck} data.} 

\revision{Previous results have also identified changes in magnetic field orientations between these scales. For example, \citet{2015Planck} link a sharp drop in the observed polarization fraction at a column density $N(\mbox{H}_2) \gtrsim 1.5 \times 10^{22}$~cm$^{-2}$ to fluctuations in the plane-of-sky B-field orientation at higher $N(\mbox{H}_2)$, rather than a true decrease in the polarization fraction at higher densities. 
This matches the median minimum column density, $N(\mbox{H}_2) \sim 2 \times 10^{22}$~cm$^{-2}$, we calculate for our JCMT data given the sensitivity thresholds in $I$ and $p$ \citep[following standard methods to derive $N(\mbox{H}_2)$ from continuum dust emission in][assuming dust temperature $T_d = 12$~K and dust opacity exponent $\beta = 1.7$]{kauffmann_2008}.  
The JCMT data are thus sensitive to only a small fraction of the emission traced along the line-of-sight by \textit{Planck}. The \textit{Planck} and JCMT datasets therefore reveal the B-field orientation at different physical scales, and also between low and high column density structures in these regions.} 

\subsection{Alignment and environment}

\revision{As discussed in Section \ref{subsec:results_field}, we classified each of the 14 star-forming regions studied into three cases based on the overall alignment between cloud- and core-scales, and how ordered the fields were based on the dispersion in the polarization vectors. Below, we discuss the broader environment and evolutionary stage and how these factors may play a role in the regions within our three classifications.} 

\subsubsection{B-field and structure alignment}

\revision{As discussed in Section \ref{sec:intro}, alignment between filamentary structure and the cloud-scale and core-scale fields varies between regions. 
This relative orientation may change, however, as a function of spatial scale, or as the region evolves. 
For example, observations have revealed changes in orientation on small scales toward a parallel orientation due to filamentary gas flows \citep{pillai_2020}, or more generally due to gravitational contraction \citep[e.g.,][]{li_2015,pattle_2017}.} 

\revision{Two of 14 regions in our study (B1 and IC348, both in the Perseus molecular cloud) are classified as Case 1, where the mean core-scale and cloud-scale magnetic field orientation aligns within 30$^\circ$ and the standard deviation of the two distributions fall under 30$^\circ$.  
B1 is highly elongated, and the magnetic fields traced by both \textit{Planck} and JCMT are aligned approximately perpendicular to the main filament \citep{GAS_perseus_filaments}. 
The dispersion in both the JCMT and \textit{Planck} vectors is low, suggesting that the magnetic field remains well-connected in the transition from cloud-scales to core-scales in B1. 
The northern component of B1 becomes more complicated, however, and some filamentary structures are aligned parallel with the field on both cloud- and core-scales.
\revision{Some simulations predict that the core-scale field should be aligned in general with the cloud-scale field \citep{Chen2020, chen_ostriker_2018}. While some regions show alignment between the core-scale and cloud-scale magnetic fields, they make up a small fraction of our observed regions, indicating that} this alignment between the core-scale and cloud-scale magnetic fields may be rare. 
Observations of IC348 are very limited spatially (see Figure \ref{fig:perseus_visual}), precluding any broad conclusions regarding the relative alignment of magnetic fields and spatial structures in this region.}

\revision{Six of the 14 observed regions are classified as Case 2, where the mean core-scale and cloud-scale magnetic field orientations align within 30$^\circ$, but with a standard deviation in the JCMT data that is higher than 30$^\circ$, indicating that the core-scale magnetic field is disordered. 
The final six regions show both misalignment between the mean core- and cloud-scale magnetic orientations, in addition to greater disorder in the core-scale orientation (Case 3). 
The physical structures in these regions are generally more complex than in, e.g., B1. 
In conjunction with the more disordered core-scale field, a comparison of the magnetic field alignment to filamentary structure is more challenging.} 

\revision{In some cases, the B-field may be more ordered on small scales than is evident from the averaged dispersion in orientation, if it is aligned with substructures in the regions rather than the cloud-scale field. For example, \cite{JCMTBISTRO_NGC1333_2020} argue that individual filaments within the NGC 1333 complex are aligned perpendicularly to the magnetic field, but show variations in the observed relative alignment due to 2D projection effects.
Within three of the eight filaments analyzed, however, the magnetic fields are still disordered, with dispersions in the orientation angles $> 30^\circ$. \citeauthor{JCMTBISTRO_NGC1333_2020} conclude that the magnetic field orientation becomes more complex at spatial scales $\lesssim 1$ pc, in line with our findings.} 

\revision{Nearby regions within the same molecular cloud can also show different relative orientations between core- and cloud-scale magnetic fields, as well as with the overall cloud structure. Toward L1689, for example, L1689-1 falls in our Case 2 category, while the nearby L1689-2 is Case 3. \cite{BISTRO_L1689} show that while the JCMT and \textit{Planck} fields are both aligned perpendicular to the filament in L1689-1 (their L1689N), the \textit{Planck} field is aligned parallel to the filament in L1689-2 (their SMM-16) while the JCMT field is aligned perpendicular. 
Similarly, the filamentary structure of NGC 2024 in Orion B is aligned approximately parallel to the \textit{Planck} field and perpendicular to the JCMT field\revision{, though the JCMT field only covers a small part of the region. Data from the HAWC+ instrument aboard SOFIA aligns well with the JCMT field in the regions of NGC 2024 where the observations overlap. In some lower column density regions of NGC 2024 outside of the JCMT observations, however, the HAWC+ field aligns with the \textit{Planck} field \citep{beslic_2024}.}
In contrast, a similar perpendicular alignment between the cloud- and core-scale magnetic field orientations is seen in L1455, but a qualitative examination of the filament structure \citep{GAS_perseus_filaments} reveals that it is the core-scale field that is aligned more parallel to the filament, while the cloud-scale field is aligned perpendicular to the filament.}

\revision{The relative orientation between the magnetic field and filamentary structure is also seen to vary within an individual region. For example, toward the highly filamentary OMC 2/3 region, on cloud scales the field appears largely perpendicular to the filament spine in the north, but becomes more disordered in the south toward OMC-1 \citep[Figure \ref{fig:orion_visual}; see also][]{Soler2019}. On core scales, the orientation of the field appears to vary relative to the filament. \citet{poidevin_2010} argue that some regions show parallel and some show perpendicular orientations between the filament spine and the magnetic field vectors. \citet{jiao_2024} claim a similar bimodal distribution of magnetic field orientations toward OMC 2/3, but find also that the distribution is consistent with being random on small scales, in agreement with our finding of large dispersion in the core-scale magnetic field vectors.} 

\revision{Within the observed regions, we therefore find a complex pattern of relative orientation between cloud- and core-scale fields, and their physical structure. 
This suggests that there is not a consistent preferred core-scale alignment to the filamentary structure, when the core-scale and cloud-scale fields do not align. 
Some of this variation may be driven by feedback effects, which we discuss further below. 
A quantitative analysis of the relative orientation of filaments and magnetic fields is needed to probe more deeply the transition from cloud to core, and additionally identify the effects of projection in individual regions, but is beyond the scope of this paper.}  

\subsubsection{Gravitational collapse}

\revision{The increased disorder in the core-scale field in most of the regions studied here could be the result of several different factors. One possibility is that at higher gas column densities, the magnetic field gets tangled and pulled by gravity as the magnetic field becomes less dynamically important. In MHD simulations of prestellar core-forming regions, \cite{Chen2016} concluded that after a self-gravitating prestellar core is formed, the magnetic field will become increasingly tangled due to the gravity-induced isotropic contraction in the magnetic field. Furthermore, \citet{2025Tritsis} show that even when a contracting core has an ordered, hourglass-shaped magnetic field in 3D, it will often appear more tangled if not viewed edge-on. We note, however, that many of the cores in this sample are not yet gravitationally unstable, based on a virial analysis \citep{Pineda_2026_GAS_DR2}, but still show disorder in their magnetic field orientation.}

\subsubsection{Feedback}

\revision{Feedback from young stars can affect both the physical structures as well as the magnetic field orientation in star-forming regions. On large scales, for example, where feedback dominates over gravity, H{\small{\textsc{II}}} regions can push gas and magnetic field lines toward an almost tangential (to the bubble) magnetic field morphology, altering their original relative alignment \citep{tahani_2023}.} 

\revision{Four of the six regions in Case 3 (Oph A, Oph B, OMC2/OMC3, and NGC 2024) also show the largest dispersion in the magnetic field vectors measured with \textit{Planck} in our sample (see Table \ref{tab:region_stats}), indicating that the plane-of-sky magnetic field is already disordered on cloud scales. 
In these regions, the broad dispersion in magnetic field orientation on cloud scales, along with the observed misalignment of the cloud- and core-scale fields, may be driven by large-scale feedback effects distorting the magnetic field morphology. 
For example, two distinct components are evident in the plane-of-sky magnetic field toward NGC 2024, corresponding to a field associated with its filamentary structure, and material swept up by an expanding H{\small{\textsc{II}}} region \citep{matthews_2002, beslic_2024}.} 

\revision{Toward L1688 in Ophiuchus (containing Oph A, B, C, E, and F), IR polarimetry has revealed a strong, ordered component to the cloud-scale field \citep{sato_1988, kwon_2015}. \textit{Planck} measurements, however, also reveal substantial changes in B-field orientation near some of the higher column density features we identify as Case 3 regions, such as Oph A and B \citep{BISTRO_L1689}. \citet{alves_2025} argue that some of these structures in L1688 formed due to feedback from nearby massive stars, and the origin of the dispersion in B-field orientation on cloud scales may thus be similar in nature to NGC 2024.} 

\revision{Lower-mass stars do not drive H{\small{\textsc{II}}} regions, but introduce feedback in the forms of jets, outflows, and photo-dissociation regions on smaller scales. While the impact of these on the local magnetic field is not yet well understood, \citet{sharma_2025} show evidence for the reshaping of magnetic field lines by outflow feedback in the vicinity of a photo-dissociation region driven by an intermediate-mass, pre-main-sequence star. Toward NGC 1333, \cite{JCMTBISTRO_NGC1333_2020} suggest that the increased complexity of the magnetic field at core-scales could also be due to dynamical interaction of the molecular outflows from young stellar objects (YSOs), warping the magnetic field.} 

\revision{We investigate whether protostellar cores in our sample show evidence for increased magnetic field complexity relative to starless and prestellar cores by comparing the mean standard deviation of their core-scale magnetic fields, filtering for cores that have at least 10 magnetic field vectors.
In contrast to expectations, we find that the dispersion of the core-scale field within starless and prestellar cores is actually greater than that of protostellar cores, with a mean standard deviation of $30^\circ$ compared to $16^\circ$. Our sample size of cores, however, is limited, with only 40 cores with at least 10 magnetic field vectors, including 29 starless and prestellar cores, and 11 protostellar cores. A larger sample is needed to test this more definitively.}

\subsection{Alignment of the averaged magnetic field orientation in cores, core orientation, and core velocity gradient}

Next, we discuss the core-scale magnetic field averaged within individual cores, and the correlation between the averaged magnetic field orientation and other core properties.
We find no globally preferred alignment or anti-alignment between the core-scale magnetic field $\theta_B$, the core major axis $\theta_C$, and the core velocity gradient $\theta_G$. 
This builds on the findings of \cite{Pandhi2023} that found that the cloud-scale magnetic field traced by \textit{Planck} observations is generally not correlated with the same core properties listed above. Analysis of specific regions individually also does not reveal any significant correlations. 
In addition, within individual regions, the averaged core magnetic field orientations do not appear to preferentially align well with each other. This result is unsurprising given the increased dispersion of the core-scale magnetic field orientation in most regions, as discussed in Section \ref{sec:discussion_field}. 

The lack of any preferred alignment or anti-alignment with the core-scale magnetic field and core properties suggests that the magnetic field does not play a dominant role in the evolution of dense cores at core-scales, contrary to classical models and the findings of some MHD simulations \citep{chen_ostriker_2018, Chen2020}. In the classical view of star formation, dense cores are modeled as oblate spheroids rotating about their minor axis and flattened along the magnetic field direction, which is parallel to the minor axis \citep{mestel_spitzer_1956, strittmatter_1966, mouschovias_1976, crutcher_1999}. Following this, we would expect to see perpendicular alignment between both $\theta_B$ and $\theta_C$, and $\theta_B$ and $\theta_G$. Instead, we find distributions consistent with random. \cite{Lee2017} produced synthetic observations of 3D MHD simulations of low-mass cores, and found that weakly magnetized cores displayed random alignment between $\theta_B$ and $\theta_G$, which is compatible with our results. Similarly, \cite{kuznetsova2020} find random alignment between the angular momentum vector of cores and the core-scale magnetic field. In contrast, \cite{chen_ostriker_2018} find that their simulated cores are generally triaxial and the core-scale magnetic field is overall preferentially aligned perpendicular to the core major axis (which we do not find), while randomly aligned to the velocity gradient (which we do find). \cite{Chen2020} finds that in 3D space, their simulated cores have a weak but clear preferential perpendicular alignment to the local, small-scale magnetic field, which is again at odds with our results, but they find that the cores do not have a strong preferential alignment to the large-scale field, in agreement with our results. Both the \cite{chen_ostriker_2018} and \cite{Chen2020} simulations were conducted in 3D space, so it is possible we may not see as strong of an alignment from our 2D-projected data. In general, there is no one-to-one mapping between initial conditions of the simulations and the observed correlation between the magnetic field and core properties, but the simulations above that have stronger initial magnetic field strengths tend to produce more alignment between core properties and the local magnetic field.

\revision{Another mechanism that could drive disorder in the magnetic field is increased turbulent motions, which we can trace using the non-thermal component of the NH$_3$ linewidth. To test whether there is a correlation, we compare the standard deviation of magnetic field vectors assigned to each individual core with the average non-thermal linewidth within each core provided in \cite{Pineda_2026_GAS_DR2}. We find that there is no correlation between the mean core non-thermal linewidth and the standard deviation of the magnetic field vectors within the core. The cores in our sample have very little variation in the non-thermal linewidth, with a mean and standard deviation of 0.32 and 0.21 km/s respectively for all cores, and a mean and standard deviation of 0.27 and 0.11 km/s respectively when excluding cores from OMC-1, which are overall more turbulent. In general, NH$_3$ selectively traces dense gas where in general turbulent motions have dissipated. A comparison of non-thermal motions as traced by molecules like CO, which are excited in lower-density gas, may thus reveal a correlation that is not seen in our data.}

Observational results analyzing the alignment of core-scale magnetic field orientation of cores to other properties also vary. In the Taurus B213 filament, BISTRO data revealed that among three dense cores, while the core-scale magnetic field orientation of one core was consistent with the large-scale field, the two other cores were misaligned \citep{Taurus2021}. \cite{Taurus2021} suggests the misalignment in the two dense cores is caused by localized differences in gas kinematics, which may arise due to gas inflows onto the filament. Other studies have found varying degrees of alignment between dense cores and the magnetic field, from random to some degree of preferential alignment or anti-alignment \citep{Chen2020, arce-tord_2020, sharma2022}.

We do not observe the significant perpendicular alignment between $\theta_B$ and $\theta_C$ found for protostellar cores in \cite{Pandhi2023}.
It is important to note, however, that we have a much smaller sample size of protostellar cores than \cite{Pandhi2023} (31 versus their 113), given that we are using a more limited map area. Nevertheless, our analysis is performed at much higher resolution that resolves the magnetic field morphology in the cores at 21 times better resolution than \citeauthor{Pandhi2023}. The difference in our results implies that the cloud-scale magnetic field displays a preferential anti-alignment to the major axis of protostellar cores, while the core-scale magnetic field does not. \cite{Pandhi2023} concluded that their result could either indicate that alignment between $\theta_B$ and $\theta_C$ evolves from randomly aligned in starless and prestellar cores to anti-aligned for protostellar cores, or that we are observing a selection effect, where starless and prestellar cores that are anti-aligned to the cloud-scale magnetic field are more likely to evolve to become protostellar. Our finding that the core-scale magnetic field does not display this preferential anti-alignment to protostellar cores is more favorable to the latter scenario as, in the former scenario, we would expect to see some re-organizing in the core-scale magnetic field if cores were developing an anti-alignment to the cloud-scale field during their evolution into protostellar cores. 

\section{Conclusion}\label{sec:conclusion}

We use dust polarization observations from the JCMT via POL-2 and SCUPOL to calculate the core-scale magnetic field across 14 star-forming regions. Cross-matching with information from core catalogs sourced from the \textit{Herschel} Gould Belt Survey (HGBS) and the JCMT Gould Belt Survey (JCMT GBS), as well as from \cite{Pandhi2023}, we additionally produce a catalog of 79 cores with their core-scale and cloud-scale magnetic field orientations, core orientation, and velocity gradient orientation. The key conclusions of this paper are summarized below:

\begin{enumerate}
  \item In general, we find clear changes in the magnetic field in the transition from cloud- to core-scales. We argue that JCMT and \textit{Planck} are indeed tracing different field morphologies and that, overall, the core-scale magnetic field displays a much more disordered and complex morphology than the cloud-scale field.
  \item Alignment between the cloud- and core-scale magnetic field varies greatly between regions, which we categorize into three different cases of alignment: Case 1 (aligned and ordered), Case 2 (aligned and disordered), and Case 3 (misaligned) (see details of categorization in Section \ref{sec:results}).
  \begin{enumerate}
    \item Case 1: We find that only 2 of 14 total regions are classified as Case 1: B1 and IC348. The orientations of the cloud- and core-scale magnetic fields in these regions agree with simulations that predict the core-scale field is generally aligned with the cloud-scale field, as well as observations that conclude the magnetic field is generally aligned perpendicular to dense filaments. However, the small fraction of regions classified as Case 1 indicates that a strong alignment between the core-scale and cloud-scale magnetic fields may be uncommon. 
    \item Case 2: We find that 6 of 14 total regions are classified as Case 2: NGC 1333 and L1448 in Perseus, Oph C/E/F and L1689-1 in Ophiuchus, and OMC 1 and NGC 2068 in Orion A and B, respectively. The increased disorder in the core-scale field could be due to a variety of factors, including the influence of gravity, turbulence, and/or outflows. We conclude that the magnetic field is no longer dynamically dominant on core scales in these regions.
    \item Case 3: We find that 6 of 14 total region are classified as Case 3: L1455 in Perseus; Oph A, Oph B, L1689-2 in Ophiuchus; and OMC2/OMC3 and NGC 2024 in Orion A and B, respectively. The increased disorder in the core-scale field could be caused by similar factors as for Case 2 regions. However, the misalignment between the core- and cloud-scale fields, and the fact that most of these regions also display a disordered cloud-scale field, indicates that there may be external, environmental factors causing even the cloud-scale magnetic field to be less dynamically important in these regions.
  \end{enumerate}
  \item We find no globally preferred alignment between the core-scale magnetic field, core orientation, and core velocity gradient orientation. Our results suggest that the core-scale magnetic field may not play an important role in the formation/evolution of dense cores, contrary to classical models and the results of several MHD simulations.
  \item Notably, we do not find a preferential anti-alignment between the core-scale magnetic field and core orientation for protostellar cores in particular. This is in contrast to the results by \cite{Pandhi2023}, \revision{who} found such a correlation with the cloud-scale magnetic field. This implies that \revision{while} the cloud-scale magnetic field is preferentially anti-aligned to the major axis of protostellar cores, the core-scale magnetic field is not. Our data are consistent with the model where we are observing a selection effect, in which starless and prestellar cores that are anti-aligned to the cloud-scale magnetic field are more likely to evolve to become protostellar. We caution, however, that with a sample of only 31 protostellar cores, we are not able to definitively assess cause and effect in this analysis.
\end{enumerate}

\section*{Acknowledgments}
\revision{We thank the anonymous referee for their constructive comments which improved the paper.}
The authors would like to thank the Dunlap Institute for Astronomy and Astrophysics for their support through the Summer Undergraduate Research Program (SURP) and publication funding. The Dunlap Institute is funded through an endowment established by the David Dunlap family and the University of Toronto.
The University of Toronto operates on the traditional land of the Huron-Wendat, the Seneca, and most recently, the Mississaugas of the Credit River; we are grateful to have the opportunity to work on this land. 
\revision{A.P. is a Trottier Space Institute Postdoctoral Fellow.}
D.J.\ is supported by NRC Canada and by an NSERC Discovery Grant.
F.P. acknowledges support from the Spanish Ministerio de Ciencia, Innovaci\'on y Universidades (MICINN) under grant numbers PID2022-141915NB-C21.

\software{Astropy \citep{astropy:2013, astropy:2018, astropy:2022},
SciPy \citep{2020SciPy-NMeth}.}

\bibliography{citations}{}

\begin{thebibliography}{}
\expandafter\ifx\csname natexlab\endcsname\relax\def\natexlab#1{#1}\fi
\providecommand{\url}[1]{\href{#1}{#1}}
\providecommand{\dodoi}[1]{doi:~\href{http://doi.org/#1}{\nolinkurl{#1}}}
\providecommand{\doeprint}[1]{\href{http://ascl.net/#1}{\nolinkurl{http://ascl.net/#1}}}
\providecommand{\doarXiv}[1]{\href{https://arxiv.org/abs/#1}{\nolinkurl{https://arxiv.org/abs/#1}}}

\bibitem[{{Alves} {et~al.}(2025){Alves}, {Lombardi}, \& {Lada}}]{alves_2025}
{Alves}, J., {Lombardi}, M., \& {Lada}, C. 2025, arXiv e-prints, arXiv:2501.13931, \dodoi{10.48550/arXiv.2501.13931}

\bibitem[{Anderson \& Darling(1954)}]{ad_test}
Anderson, T.~W., \& Darling, D.~A. 1954, Journal of the American Statistical Association, 49, 765, \dodoi{10.1080/01621459.1954.10501232}

\bibitem[{{Andersson} {et~al.}(2015){Andersson}, {Lazarian}, \& {Vaillancourt}}]{Andersson_2015}
{Andersson}, B.~G., {Lazarian}, A., \& {Vaillancourt}, J.~E. 2015, \araa, 53, 501, \dodoi{10.1146/annurev-astro-082214-122414}

\bibitem[{{Andr{\'e}} {et~al.}(2010){Andr{\'e}}, {Men'shchikov}, {Bontemps}, {K{\"o}nyves}, {Motte}, {Schneider}, {Didelon}, {Minier}, {Saraceno}, {Ward-Thompson}, {di Francesco}, {White}, {Molinari}, {Testi}, {Abergel}, {Griffin}, {Henning}, {Royer}, {Mer{\'\i}n}, {Vavrek}, {Attard}, {Arzoumanian}, {Wilson}, {Ade}, {Aussel}, {Baluteau}, {Benedettini}, {Bernard}, {Blommaert}, {Cambr{\'e}sy}, {Cox}, {di Giorgio}, {Hargrave}, {Hennemann}, {Huang}, {Kirk}, {Krause}, {Launhardt}, {Leeks}, {Le Pennec}, {Li}, {Martin}, {Maury}, {Olofsson}, {Omont}, {Peretto}, {Pezzuto}, {Prusti}, {Roussel}, {Russeil}, {Sauvage}, {Sibthorpe}, {Sicilia-Aguilar}, {Spinoglio}, {Waelkens}, {Woodcraft}, \& {Zavagno}}]{HGBS_Andre}
{Andr{\'e}}, P., {Men'shchikov}, A., {Bontemps}, S., {et~al.} 2010, \aap, 518, L102, \dodoi{10.1051/0004-6361/201014666}

\bibitem[{{Arce-Tord} {et~al.}(2020){Arce-Tord}, {Louvet}, {Cortes}, {Motte}, {Hull}, {Le Gouellec}, {Garay}, {Nony}, {Didelon}, \& {Bronfman}}]{arce-tord_2020}
{Arce-Tord}, C., {Louvet}, F., {Cortes}, P.~C., {et~al.} 2020, \aap, 640, A111, \dodoi{10.1051/0004-6361/202038024}

\bibitem[{{Astropy Collaboration} {et~al.}(2013){Astropy Collaboration}, {Robitaille}, {Tollerud}, {Greenfield}, {Droettboom}, {Bray}, {Aldcroft}, {Davis}, {Ginsburg}, {Price-Whelan}, {Kerzendorf}, {Conley}, {Crighton}, {Barbary}, {Muna}, {Ferguson}, {Grollier}, {Parikh}, {Nair}, {Unther}, {Deil}, {Woillez}, {Conseil}, {Kramer}, {Turner}, {Singer}, {Fox}, {Weaver}, {Zabalza}, {Edwards}, {Azalee Bostroem}, {Burke}, {Casey}, {Crawford}, {Dencheva}, {Ely}, {Jenness}, {Labrie}, {Lim}, {Pierfederici}, {Pontzen}, {Ptak}, {Refsdal}, {Servillat}, \& {Streicher}}]{astropy:2013}
{Astropy Collaboration}, {Robitaille}, T.~P., {Tollerud}, E.~J., {et~al.} 2013, \aap, 558, A33, \dodoi{10.1051/0004-6361/201322068}

\bibitem[{{Astropy Collaboration} {et~al.}(2018){Astropy Collaboration}, {Price-Whelan}, {Sip{\H{o}}cz}, {G{\"u}nther}, {Lim}, {Crawford}, {Conseil}, {Shupe}, {Craig}, {Dencheva}, {Ginsburg}, {Vand erPlas}, {Bradley}, {P{\'e}rez-Su{\'a}rez}, {de Val-Borro}, {Aldcroft}, {Cruz}, {Robitaille}, {Tollerud}, {Ardelean}, {Babej}, {Bach}, {Bachetti}, {Bakanov}, {Bamford}, {Barentsen}, {Barmby}, {Baumbach}, {Berry}, {Biscani}, {Boquien}, {Bostroem}, {Bouma}, {Brammer}, {Bray}, {Breytenbach}, {Buddelmeijer}, {Burke}, {Calderone}, {Cano Rodr{\'\i}guez}, {Cara}, {Cardoso}, {Cheedella}, {Copin}, {Corrales}, {Crichton}, {D'Avella}, {Deil}, {Depagne}, {Dietrich}, {Donath}, {Droettboom}, {Earl}, {Erben}, {Fabbro}, {Ferreira}, {Finethy}, {Fox}, {Garrison}, {Gibbons}, {Goldstein}, {Gommers}, {Greco}, {Greenfield}, {Groener}, {Grollier}, {Hagen}, {Hirst}, {Homeier}, {Horton}, {Hosseinzadeh}, {Hu}, {Hunkeler}, {Ivezi{\'c}}, {Jain}, {Jenness}, {Kanarek}, {Kendrew}, {Kern}, {Kerzendorf}, {Khvalko}, {King}, {Kirkby}, {Kulkarni},
  {Kumar}, {Lee}, {Lenz}, {Littlefair}, {Ma}, {Macleod}, {Mastropietro}, {McCully}, {Montagnac}, {Morris}, {Mueller}, {Mumford}, {Muna}, {Murphy}, {Nelson}, {Nguyen}, {Ninan}, {N{\"o}the}, {Ogaz}, {Oh}, {Parejko}, {Parley}, {Pascual}, {Patil}, {Patil}, {Plunkett}, {Prochaska}, {Rastogi}, {Reddy Janga}, {Sabater}, {Sakurikar}, {Seifert}, {Sherbert}, {Sherwood-Taylor}, {Shih}, {Sick}, {Silbiger}, {Singanamalla}, {Singer}, {Sladen}, {Sooley}, {Sornarajah}, {Streicher}, {Teuben}, {Thomas}, {Tremblay}, {Turner}, {Terr{\'o}n}, {van Kerkwijk}, {de la Vega}, {Watkins}, {Weaver}, {Whitmore}, {Woillez}, {Zabalza}, \& {Astropy Contributors}}]{astropy:2018}
{Astropy Collaboration}, {Price-Whelan}, A.~M., {Sip{\H{o}}cz}, B.~M., {et~al.} 2018, \aj, 156, 123, \dodoi{10.3847/1538-3881/aabc4f}

\bibitem[{{Astropy Collaboration} {et~al.}(2022){Astropy Collaboration}, {Price-Whelan}, {Lim}, {Earl}, {Starkman}, {Bradley}, {Shupe}, {Patil}, {Corrales}, {Brasseur}, {N{"o}the}, {Donath}, {Tollerud}, {Morris}, {Ginsburg}, {Vaher}, {Weaver}, {Tocknell}, {Jamieson}, {van Kerkwijk}, {Robitaille}, {Merry}, {Bachetti}, {G{"u}nther}, {Aldcroft}, {Alvarado-Montes}, {Archibald}, {B{'o}di}, {Bapat}, {Barentsen}, {Baz{'a}n}, {Biswas}, {Boquien}, {Burke}, {Cara}, {Cara}, {Conroy}, {Conseil}, {Craig}, {Cross}, {Cruz}, {D'Eugenio}, {Dencheva}, {Devillepoix}, {Dietrich}, {Eigenbrot}, {Erben}, {Ferreira}, {Foreman-Mackey}, {Fox}, {Freij}, {Garg}, {Geda}, {Glattly}, {Gondhalekar}, {Gordon}, {Grant}, {Greenfield}, {Groener}, {Guest}, {Gurovich}, {Handberg}, {Hart}, {Hatfield-Dodds}, {Homeier}, {Hosseinzadeh}, {Jenness}, {Jones}, {Joseph}, {Kalmbach}, {Karamehmetoglu}, {Ka{l}uszy{'n}ski}, {Kelley}, {Kern}, {Kerzendorf}, {Koch}, {Kulumani}, {Lee}, {Ly}, {Ma}, {MacBride}, {Maljaars}, {Muna}, {Murphy}, {Norman}, {O'Steen},
  {Oman}, {Pacifici}, {Pascual}, {Pascual-Granado}, {Patil}, {Perren}, {Pickering}, {Rastogi}, {Roulston}, {Ryan}, {Rykoff}, {Sabater}, {Sakurikar}, {Salgado}, {Sanghi}, {Saunders}, {Savchenko}, {Schwardt}, {Seifert-Eckert}, {Shih}, {Jain}, {Shukla}, {Sick}, {Simpson}, {Singanamalla}, {Singer}, {Singhal}, {Sinha}, {Sip{H{o}}cz}, {Spitler}, {Stansby}, {Streicher}, {{{S}}umak}, {Swinbank}, {Taranu}, {Tewary}, {Tremblay}, {Val-Borro}, {Van Kooten}, {Vasovi{'c}}, {Verma}, {de Miranda Cardoso}, {Williams}, {Wilson}, {Winkel}, {Wood-Vasey}, {Xue}, {Yoachim}, {Zhang}, {Zonca}, \& {Astropy Project Contributors}}]{astropy:2022}
{Astropy Collaboration}, {Price-Whelan}, A.~M., {Lim}, P.~L., {et~al.} 2022, \apj, 935, 167, \dodoi{10.3847/1538-4357/ac7c74}

\bibitem[{{Ballesteros-Paredes} {et~al.}(2007){Ballesteros-Paredes}, {Klessen}, {Mac Low}, \& {Vazquez-Semadeni}}]{Ballesteros-Paredes2007}
{Ballesteros-Paredes}, J., {Klessen}, R.~S., {Mac Low}, M.~M., \& {Vazquez-Semadeni}, E. 2007, in Protostars and Planets V, ed. B.~{Reipurth}, D.~{Jewitt}, \& K.~{Keil}, 63, \dodoi{10.48550/arXiv.astro-ph/0603357}

\bibitem[{{Bastien} {et~al.}(2011){Bastien}, {Bissonnette}, {Simon}, {Coud{\'e}}, {Ade}, {Savini}, {Pisano}, {Leclerc}, {Bernier}, {Landry}, {Houde}, {Hezareh}, {Naylor}, {Gom}, {Jenness}, {Berry}, {Johnstone}, \& {Matthews}}]{POL-2_Bastien2011}
{Bastien}, P., {Bissonnette}, E., {Simon}, A., {et~al.} 2011, in Astronomical Society of the Pacific Conference Series, Vol. 449, Astronomical Polarimetry 2008: Science from Small to Large Telescopes, ed. P.~{Bastien}, N.~{Manset}, D.~P. {Clemens}, \& N.~{St-Louis}, 68

\bibitem[{{Basu}(2000)}]{2000Basu}
{Basu}, S. 2000, \apjl, 540, L103, \dodoi{10.1086/312885}

\bibitem[{{Be{\v{s}}li{\'c}} {et~al.}(2024){Be{\v{s}}li{\'c}}, {Coud{\'e}}, {Lis}, {Gerin}, {Goldsmith}, {Pety}, {Roueff}, {Demyk}, {Dowell}, {Einig}, {Goicoechea}, {Levrier}, {Orkisz}, {Peretto}, {Santa-Maria}, {Ysard}, \& {Zakardjian}}]{beslic_2024}
{Be{\v{s}}li{\'c}}, I., {Coud{\'e}}, S., {Lis}, D.~C., {et~al.} 2024, \aap, 684, A212, \dodoi{10.1051/0004-6361/202348376}

\bibitem[{{Burkert} \& {Bodenheimer}(2000)}]{2000Burkert}
{Burkert}, A., \& {Bodenheimer}, P. 2000, \apj, 543, 822, \dodoi{10.1086/317122}

\bibitem[{{Chapin} {et~al.}(2013){Chapin}, {Berry}, {Gibb}, {Jenness}, {Scott}, {Tilanus}, {Economou}, \& {Holland}}]{2013Chapin}
{Chapin}, E.~L., {Berry}, D.~S., {Gibb}, A.~G., {et~al.} 2013, \mnras, 430, 2545, \dodoi{10.1093/mnras/stt052}

\bibitem[{{Chen} {et~al.}(2016){Chen}, {King}, \& {Li}}]{Chen2016}
{Chen}, C.-Y., {King}, P.~K., \& {Li}, Z.-Y. 2016, \apj, 829, 84, \dodoi{10.3847/0004-637X/829/2/84}

\bibitem[{{Chen} \& {Ostriker}(2018)}]{chen_ostriker_2018}
{Chen}, C.-Y., \& {Ostriker}, E.~C. 2018, \apj, 865, 34, \dodoi{10.3847/1538-4357/aad905}

\bibitem[{{Chen} {et~al.}(2020){Chen}, {Behrens}, {Washington}, {Fissel}, {Friesen}, {Li}, {Pineda}, {Ginsburg}, {Kirk}, {Scibelli}, {Alves}, {Redaelli}, {Caselli}, {Punanova}, {Di Francesco}, {Rosolowsky}, {Offner}, {Martin}, {Chac{\'o}n-Tanarro}, {Chen}, {Chen}, {Keown}, {Seo}, {Shirley}, {Arce}, {Goodman}, {Matzner}, {Myers}, \& {Singh}}]{Chen2020}
{Chen}, C.-Y., {Behrens}, E.~A., {Washington}, J.~E., {et~al.} 2020, \mnras, 494, 1971, \dodoi{10.1093/mnras/staa835}

\bibitem[{{Chen} {et~al.}(2024){Chen}, {Di Francesco}, {Friesen}, {Pineda}, {Caselli}, {Ginsburg}, {Kirk}, {Punanova}, \& {The GAS Collaboration}}]{GAS_perseus_filaments}
{Chen}, M. C.-Y., {Di Francesco}, J., {Friesen}, R.~K., {et~al.} 2024, \apj, 977, 135, \dodoi{10.3847/1538-4357/ad88e8}

\bibitem[{{Coud{\'e}} {et~al.}(2019){Coud{\'e}}, {Bastien}, {Houde}, {Sadavoy}, {Friesen}, {Di Francesco}, {Johnstone}, {Mairs}, {Hasegawa}, {Kwon}, {Lai}, {Qiu}, {Ward-Thompson}, {Berry}, {Chen}, {Fiege}, {Franzmann}, {Hatchell}, {Lacaille}, {Matthews}, {Moriarty-Schieven}, {Pon}, {Andr{\'e}}, {Arzoumanian}, {Aso}, {Byun}, {Eswaraiah}, {Chen}, {Chen}, {Ching}, {Cho}, {Choi}, {Chrysostomou}, {Chung}, {Doi}, {Drabek-Maunder}, {Dowell}, {Eyres}, {Falle}, {Friberg}, {Fuller}, {Furuya}, {Gledhill}, {Graves}, {Greaves}, {Griffin}, {Gu}, {Hayashi}, {Hoang}, {Holland}, {Inoue}, {Inutsuka}, {Iwasaki}, {Jeong}, {Kanamori}, {Kataoka}, {Kang}, {Kang}, {Kang}, {Kawabata}, {Kemper}, {Kim}, {Kim}, {Kim}, {Kim}, {Kim}, {Kim}, {Kirk}, {Kobayashi}, {Koch}, {Kwon}, {Lee}, {Lee}, {Lee}, {Li}, {Li}, {Li}, {Liu}, {Liu}, {Liu}, {Liu}, {van Loo}, {Lyo}, {Matsumura}, {Nagata}, {Nakamura}, {Nakanishi}, {Ohashi}, {Onaka}, {Parsons}, {Pattle}, {Peretto}, {Pyo}, {Qian}, {Rao}, {Rawlings}, {Retter}, {Richer}, {Rigby}, {Robitaille},
  {Saito}, {Savini}, {Scaife}, {Seta}, {Shinnaga}, {Soam}, {Tamura}, {Tang}, {Tomisaka}, {Tsukamoto}, {Wang}, {Wang}, {Whitworth}, {Yen}, {Yoo}, {Yuan}, {Zenko}, {Zhang}, {Zhang}, {Zhou}, {Zhu}, \& {B-fields In STar-forming Regions Observations (BISTRO Collaboration}}]{JCMTBISTRO_B1_2019}
{Coud{\'e}}, S., {Bastien}, P., {Houde}, M., {et~al.} 2019, \apj, 877, 88, \dodoi{10.3847/1538-4357/ab1b23}

\bibitem[{{Crutcher}(1999)}]{crutcher_1999}
{Crutcher}, R.~M. 1999, \apj, 520, 706, \dodoi{10.1086/307483}

\bibitem[{{Crutcher} {et~al.}(2010){Crutcher}, {Wandelt}, {Heiles}, {Falgarone}, \& {Troland}}]{2010_Crutcher}
{Crutcher}, R.~M., {Wandelt}, B., {Heiles}, C., {Falgarone}, E., \& {Troland}, T.~H. 2010, \apj, 725, 466, \dodoi{10.1088/0004-637X/725/1/466}

\bibitem[{{Dempsey} {et~al.}(2013){Dempsey}, {Friberg}, {Jenness}, {Tilanus}, {Thomas}, {Holland}, {Bintley}, {Berry}, {Chapin}, {Chrysostomou}, {Davis}, {Gibb}, {Parsons}, \& {Robson}}]{JCMTGBS_Dempsey}
{Dempsey}, J.~T., {Friberg}, P., {Jenness}, T., {et~al.} 2013, \mnras, 430, 2534, \dodoi{10.1093/mnras/stt090}

\bibitem[{{Doi} {et~al.}(2020){Doi}, {Hasegawa}, {Furuya}, {Coud{\'e}}, {Hull}, {Arzoumanian}, {Bastien}, {Chen}, {Di Francesco}, {Friesen}, {Houde}, {Inutsuka}, {Mairs}, {Matsumura}, {Onaka}, {Sadavoy}, {Shimajiri}, {Tahani}, {Tomisaka}, {Eswaraiah}, {Koch}, {Pattle}, {Won Lee}, {Tamura}, {Berry}, {Ching}, {Hwang}, {Kwon}, {Soam}, {Wang}, {Lai}, {Qiu}, {Ward-Thompson}, {Byun}, {Chen}, {Chen}, {Chen}, {Cho}, {Choi}, {Choi}, {Chrysostomou}, {Chung}, {Diep}, {Duan}, {Fanciullo}, {Fiege}, {Franzmann}, {Friberg}, {Fuller}, {Gledhill}, {Graves}, {Greaves}, {Griffin}, {Gu}, {Han}, {Hatchell}, {Hayashi}, {Hoang}, {Inoue}, {Iwasaki}, {Jeong}, {Johnstone}, {Kanamori}, {Kang}, {Kang}, {Kang}, {Kataoka}, {Kawabata}, {Kemper}, {Kim}, {Kim}, {Kim}, {Kim}, {Kim}, {Kim}, {Kirk}, {Kobayashi}, {Konyves}, {Kusune}, {Kwon}, {Lacaille}, {Law}, {Lee}, {Lee}, {Lee}, {Lee}, {Lee}, {Li}, {Li}, {Li}, {Liu}, {Liu}, {Liu}, {Liu}, {de Looze}, {Lyo}, {Matthews}, {Moriarty-Schieven}, {Nagata}, {Nakamura}, {Nakanishi}, {Ohashi}, {Park},
  {Parsons}, {Peretto}, {Pyo}, {Qian}, {Rao}, {Rawlings}, {Retter}, {Richer}, {Rigby}, {Saito}, {Savini}, {Scaife}, {Seta}, {Shinnaga}, {Tang}, {Tsukamoto}, {Viti}, {Wang}, {Whitworth}, {Yen}, {Yoo}, {Yuan}, {Yun}, {Zenko}, {Zhang}, {Zhang}, {Zhang}, {Zhou}, {Zhu}, {Andr{\'e}}, {Dowell}, {Eyres}, {Falle}, {van Loo}, \& {Robitaille}}]{JCMTBISTRO_NGC1333_2020}
{Doi}, Y., {Hasegawa}, T., {Furuya}, R.~S., {et~al.} 2020, \apj, 899, 28, \dodoi{10.3847/1538-4357/aba1e2}

\bibitem[{{Eswaraiah} {et~al.}(2021){Eswaraiah}, {Li}, {Furuya}, {Hasegawa}, {Ward-Thompson}, {Qiu}, {Ohashi}, {Pattle}, {Sadavoy}, {Hull}, {Berry}, {Doi}, {Ching}, {Lai}, {Wang}, {Koch}, {Kwon}, {Kwon}, {Bastien}, {Arzoumanian}, {Coud{\'e}}, {Soam}, {Fanciullo}, {Yen}, {Liu}, {Hoang}, {Ping Chen}, {Shimajiri}, {Liu}, {Chen}, {Li}, {Lyo}, {Hwang}, {Johnstone}, {Rao}, {Bich Ngoc}, {Ngoc Diep}, {Mairs}, {Parsons}, {Tamura}, {Tahani}, {Vivien Chen}, {Nakamura}, {Shinnaga}, {Tang}, {Cho}, {Won Lee}, {Inutsuka}, {Inoue}, {Iwasaki}, {Qian}, {Xie}, {Li}, {Liu}, {Zhang}, {Chen}, {Zhang}, {Zhu}, {Zhou}, {Andr{\'e}}, {Liu}, {Yuan}, {Lu}, {Peretto}, {Bourke}, {Byun}, {Dai}, {Duan}, {Duan}, {Eden}, {Matthews}, {Fiege}, {Fissel}, {Kim}, {Lee}, {Kim}, {Pyo}, {Choi}, {Choi}, {Chrysostomou}, {Jung Chung}, {Ngoc Tram}, {Franzmann}, {Friberg}, {Friesen}, {Fuller}, {Gledhill}, {Graves}, {Greaves}, {Griffin}, {Gu}, {Han}, {Hatchell}, {Hayashi}, {Houde}, {Kawabata}, {Jeong}, {Kang}, {Kang}, {Kang}, {Kataoka}, {Kemper},
  {Rawlings}, {Rawlings}, {Retter}, {Richer}, {Rigby}, {Saito}, {Savini}, {Scaife}, {Seta}, {Kim}, {Hee Kim}, {Kim}, {Kirchschlager}, {Kirk}, {Kobayashi}, {Konyves}, {Kusune}, {Lacaille}, {Law}, {Lee}, {Lee}, {Matsumura}, {Moriarty-Schieven}, {Nagata}, {Nakanishi}, {Onaka}, {Park}, {Tang}, {Tomisaka}, {Tsukamoto}, {Viti}, {Wang}, {Whitworth}, {Yoo}, {Yun}, {Zenko}, {Zhang}, {de Looze}, {Dowell}, {Eyres}, {Falle}, {Robitaille}, \& {van Loo}}]{Taurus2021}
{Eswaraiah}, C., {Li}, D., {Furuya}, R.~S., {et~al.} 2021, \apjl, 912, L27, \dodoi{10.3847/2041-8213/abeb1c}

\bibitem[{{Fissel} {et~al.}(2016){Fissel}, {Ade}, {Angil{\`e}}, {Ashton}, {Benton}, {Devlin}, {Dober}, {Fukui}, {Galitzki}, {Gandilo}, {Klein}, {Korotkov}, {Li}, {Martin}, {Matthews}, {Moncelsi}, {Nakamura}, {Netterfield}, {Novak}, {Pascale}, {Poidevin}, {Santos}, {Savini}, {Scott}, {Shariff}, {Diego Soler}, {Thomas}, {Tucker}, {Tucker}, \& {Ward-Thompson}}]{2016Fissel}
{Fissel}, L.~M., {Ade}, P. A.~R., {Angil{\`e}}, F.~E., {et~al.} 2016, \apj, 824, 134, \dodoi{10.3847/0004-637X/824/2/134}

\bibitem[{{Friberg} {et~al.}(2016){Friberg}, {Bastien}, {Berry}, {Savini}, {Graves}, \& {Pattle}}]{POL-2_Friberg2016}
{Friberg}, P., {Bastien}, P., {Berry}, D., {et~al.} 2016, in Society of Photo-Optical Instrumentation Engineers (SPIE) Conference Series, Vol. 9914, Millimeter, Submillimeter, and Far-Infrared Detectors and Instrumentation for Astronomy VIII, ed. W.~S. {Holland} \& J.~{Zmuidzinas}, 991403, \dodoi{10.1117/12.2231943}

\bibitem[{{Friesen} {et~al.}(2017){Friesen}, {Pineda}, {co-PIs}, {Rosolowsky}, {Alves}, {Chac{\'o}n-Tanarro}, {How-Huan Chen}, {Chun-Yuan Chen}, {Di Francesco}, {Keown}, {Kirk}, {Punanova}, {Seo}, {Shirley}, {Ginsburg}, {Hall}, {Offner}, {Singh}, {Arce}, {Caselli}, {Goodman}, {Martin}, {Matzner}, {Myers}, {Redaelli}, \& {GAS Collaboration}}]{GAS_Friesen}
{Friesen}, R.~K., {Pineda}, J.~E., {co-PIs}, {et~al.} 2017, \apj, 843, 63, \dodoi{10.3847/1538-4357/aa6d58}

\bibitem[{{Girart} {et~al.}(2006){Girart}, {Rao}, \& {Marrone}}]{2006Girart}
{Girart}, J.~M., {Rao}, R., \& {Marrone}, D.~P. 2006, Science, 313, 812, \dodoi{10.1126/science.1129093}

\bibitem[{{Greaves} {et~al.}(2003){Greaves}, {Holland}, {Jenness}, {Chrysostomou}, {Berry}, {Murray}, {Tamura}, {Robson}, {Ade}, {Nartallo}, {Stevens}, {Momose}, {Morino}, {Moriarty-Schieven}, {Gannaway}, \& {Haynes}}]{SCUPOL_Greaves}
{Greaves}, J.~S., {Holland}, W.~S., {Jenness}, T., {et~al.} 2003, \mnras, 340, 353, \dodoi{10.1046/j.1365-8711.2003.06230.x}

\bibitem[{{Hwang} {et~al.}(2021){Hwang}, {Kim}, {Pattle}, {Kwon}, {Sadavoy}, {Koch}, {Hull}, {Johnstone}, {Furuya}, {Won Lee}, {Arzoumanian}, {Tahani}, {Eswaraiah}, {Liu}, {Kirchschlager}, {Kim}, {Tamura}, {Kwon}, {Lyo}, {Soam}, {Kang}, {Bourke}, {Matsumura}, {Mairs}, {Kim}, {Park}, {Nakamura}, {Onaka}, {Tang}, {Liu}, {Ward-Thompson}, {Li}, {Hoang}, {Hasegawa}, {Qiu}, {Lai}, \& {Bastien}}]{JCMTBISTRO_OMC1_2021}
{Hwang}, J., {Kim}, J., {Pattle}, K., {et~al.} 2021, \apj, 913, 85, \dodoi{10.3847/1538-4357/abf3c4}

\bibitem[{{Jiao} {et~al.}(2024){Jiao}, {Wang}, {Xu}, {Wang}, \& {Beuther}}]{jiao_2024}
{Jiao}, W., {Wang}, K., {Xu}, F., {Wang}, C., \& {Beuther}, H. 2024, \aap, 686, A202, \dodoi{10.1051/0004-6361/202449182}

\bibitem[{{Kauffmann} {et~al.}(2008){Kauffmann}, {Bertoldi}, {Bourke}, {Evans}, \& {Lee}}]{kauffmann_2008}
{Kauffmann}, J., {Bertoldi}, F., {Bourke}, T.~L., {Evans}, II, N.~J., \& {Lee}, C.~W. 2008, \aap, 487, 993, \dodoi{10.1051/0004-6361:200809481}

\bibitem[{Klotz(1967)}]{ks_test}
Klotz, J. 1967, Journal of the American Statistical Association, 62, 932, \dodoi{10.1080/01621459.1967.10500904}

\bibitem[{Kolmogorov(1933)}]{Kolmogorov1933}
Kolmogorov, A.~N. 1933, Foundations of the Theory of Probability

\bibitem[{{K{\"o}nyves} {et~al.}(2020){K{\"o}nyves}, {Andr{\'e}}, {Arzoumanian}, {Schneider}, {Men'shchikov}, {Bontemps}, {Ladjelate}, {Didelon}, {Pezzuto}, {Benedettini}, {Bracco}, {Di Francesco}, {Goodwin}, {Rygl}, {Shimajiri}, {Spinoglio}, {Ward-Thompson}, \& {White}}]{HGBSOrionB}
{K{\"o}nyves}, V., {Andr{\'e}}, P., {Arzoumanian}, D., {et~al.} 2020, \aap, 635, A34, \dodoi{10.1051/0004-6361/201834753}

\bibitem[{{Kuznetsova} {et~al.}(2020){Kuznetsova}, {Hartmann}, \& {Heitsch}}]{kuznetsova2020}
{Kuznetsova}, A., {Hartmann}, L., \& {Heitsch}, F. 2020, \apj, 893, 73, \dodoi{10.3847/1538-4357/ab7eac}

\bibitem[{{Kwon} {et~al.}(2015){Kwon}, {Tamura}, {Hough}, {Nakajima}, {Nishiyama}, {Kusakabe}, {Nagata}, \& {Kandori}}]{kwon_2015}
{Kwon}, J., {Tamura}, M., {Hough}, J.~H., {et~al.} 2015, \apjs, 220, 17, \dodoi{10.1088/0067-0049/220/1/17}

\bibitem[{{Kwon} {et~al.}(2018){Kwon}, {Doi}, {Tamura}, {Matsumura}, {Pattle}, {Berry}, {Sadavoy}, {Matthews}, {Ward-Thompson}, {Hasegawa}, {Furuya}, {Pon}, {Di Francesco}, {Arzoumanian}, {Hayashi}, {Kawabata}, {Onaka}, {Choi}, {Kang}, {Hoang}, {Lee}, {Lee}, {Liu}, {Liu}, {Inutsuka}, {Eswaraiah}, {Bastien}, {Kwon}, {Lai}, {Qiu}, {Coud{\'e}}, {Franzmann}, {Friberg}, {Graves}, {Greaves}, {Houde}, {Johnstone}, {Kirk}, {Koch}, {Li}, {Parsons}, {Rao}, {Rawlings}, {Shinnaga}, {van Loo}, {Aso}, {Byun}, {Chen}, {Chen}, {Chen}, {Ching}, {Cho}, {Chrysostomou}, {Chung}, {Drabek-Maunder}, {Eyres}, {Fiege}, {Friesen}, {Fuller}, {Gledhill}, {Griffin}, {Gu}, {Hatchell}, {Holland}, {Inoue}, {Iwasaki}, {Jeong}, {Kang}, {Kang}, {Kemper}, {Kim}, {Kim}, {Kim}, {Kim}, {Kim}, {Kim}, {Lacaille}, {Lee}, {Li}, {Li}, {Liu}, {Liu}, {Lyo}, {Mairs}, {Moriarty-Schieven}, {Nakamura}, {Nakanishi}, {Ohashi}, {Peretto}, {Pyo}, {Qian}, {Retter}, {Richer}, {Rigby}, {Robitaille}, {Savini}, {Scaife}, {Soam}, {Tang}, {Tomisaka}, {Wang}, {Wang},
  {Whitworth}, {Yen}, {Yoo}, {Yuan}, {Zhang}, {Zhang}, {Zhou}, {Zhu}, {Andr{\'e}}, {Dowell}, {Falle}, {Tsukamoto}, {Nakagawa}, {Kanamori}, {Kataoka}, {Kobayashi}, {Nagata}, {Saito}, {Seta}, \& {Zenko}}]{BISTRO_OphA}
{Kwon}, J., {Doi}, Y., {Tamura}, M., {et~al.} 2018, \apj, 859, 4, \dodoi{10.3847/1538-4357/aabd82}

\bibitem[{{Ladjelate} {et~al.}(2020){Ladjelate}, {Andr{\'e}}, {K{\"o}nyves}, {Ward-Thompson}, {Men'shchikov}, {Bracco}, {Palmeirim}, {Roy}, {Shimajiri}, {Kirk}, {Arzoumanian}, {Benedettini}, {Di Francesco}, {Fiorellino}, {Schneider}, {Pezzuto}, {Motte}, \& {Herschel Gould Belt Survey Team}}]{HGBSOphiuchus}
{Ladjelate}, B., {Andr{\'e}}, P., {K{\"o}nyves}, V., {et~al.} 2020, \aap, 638, A74, \dodoi{10.1051/0004-6361/201936442}

\bibitem[{{Lazarian}(2007)}]{Lazarian2007}
{Lazarian}, A. 2007, \jqsrt, 106, 225, \dodoi{10.1016/j.jqsrt.2007.01.038}

\bibitem[{{Lazarian} \& {Hoang}(2007)}]{2007Lazarian_2}
{Lazarian}, A., \& {Hoang}, T. 2007, \mnras, 378, 910, \dodoi{10.1111/j.1365-2966.2007.11817.x}

\bibitem[{{Lee} {et~al.}(2017){Lee}, {Hull}, \& {Offner}}]{Lee2017}
{Lee}, J. W.~Y., {Hull}, C. L.~H., \& {Offner}, S. S.~R. 2017, \apj, 834, 201, \dodoi{10.3847/1538-4357/834/2/201}

\bibitem[{{Li} {et~al.}(2015){Li}, {Yuen}, {Otto}, {Leung}, {Sridharan}, {Zhang}, {Liu}, {Tang}, \& {Qiu}}]{li_2015}
{Li}, H.-B., {Yuen}, K.~H., {Otto}, F., {et~al.} 2015, \nat, 520, 518, \dodoi{10.1038/nature14291}

\bibitem[{{Liu} {et~al.}(2019){Liu}, {Qiu}, {Berry}, {Di Francesco}, {Bastien}, {Koch}, {Furuya}, {Kim}, {Coud{\'e}}, {Lee}, {Soam}, {Eswaraiah}, {Li}, {Hwang}, {Lyo}, {Pattle}, {Hasegawa}, {Kwon}, {Lai}, {Ward-Thompson}, {Ching}, {Chen}, {Gu}, {Li}, {Li}, {Liu}, {Qian}, {Wang}, {Yuan}, {Zhang}, {Zhang}, {Zhang}, {Zhou}, {Zhu}, {Andr{\'e}}, {Arzoumanian}, {Aso}, {Byun}, {Chen}, {Chen}, {Chen}, {Cho}, {Choi}, {Chrysostomou}, {Chung}, {Doi}, {Drabek-Maunder}, {Dowell}, {Eyres}, {Falle}, {Fanciullo}, {Fiege}, {Franzmann}, {Friberg}, {Friesen}, {Fuller}, {Gledhill}, {Graves}, {Greaves}, {Griffin}, {Han}, {Hatchell}, {Hayashi}, {Hoang}, {Holland}, {Houde}, {Inoue}, {Inutsuka}, {Iwasaki}, {Jeong}, {Johnstone}, {Kanamori}, {Kang}, {Kang}, {Kang}, {Kataoka}, {Kawabata}, {Kemper}, {Kim}, {Kim}, {Kim}, {Kim}, {Kim}, {Kirk}, {Kobayashi}, {Kusune}, {Kwon}, {Lacaille}, {Lee}, {Lee}, {Lee}, {Lee}, {Liu}, {Liu}, {van Loo}, {Mairs}, {Matsumura}, {Matthews}, {Moriarty-Schieven}, {Nagata}, {Nakamura}, {Nakanishi}, {Ohashi},
  {Onaka}, {Parker}, {Parsons}, {Pascale}, {Peretto}, {Pon}, {Pyo}, {Rao}, {Rawlings}, {Retter}, {Richer}, {Rigby}, {Robitaille}, {Sadavoy}, {Saito}, {Savini}, {Scaife}, {Seta}, {Shinnaga}, {Tamura}, {Tang}, {Tomisaka}, {Tsukamoto}, {Wang}, {Whitworth}, {Yen}, {Yoo}, \& {Zenko}}]{BISTRO_OphC}
{Liu}, J., {Qiu}, K., {Berry}, D., {et~al.} 2019, \apj, 877, 43, \dodoi{10.3847/1538-4357/ab0958}

\bibitem[{{Matthews} {et~al.}(2009){Matthews}, {McPhee}, {Fissel}, \& {Curran}}]{2009SCUPOL}
{Matthews}, B.~C., {McPhee}, C.~A., {Fissel}, L.~M., \& {Curran}, R.~L. 2009, \apjs, 182, 143, \dodoi{10.1088/0067-0049/182/1/143}

\bibitem[{{Matthews} \& {Wilson}(2002)}]{matthews_2002}
{Matthews}, B.~C., \& {Wilson}, C.~D. 2002, \apj, 571, 356, \dodoi{10.1086/339915}

\bibitem[{{Men'shchikov}(2021)}]{getsf}
{Men'shchikov}, A. 2021, \aap, 654, A78, \dodoi{10.1051/0004-6361/202141533}

\bibitem[{{Men'shchikov} {et~al.}(2012){Men'shchikov}, {Andr{\'e}}, {Didelon}, {Motte}, {Hennemann}, \& {Schneider}}]{getsources}
{Men'shchikov}, A., {Andr{\'e}}, P., {Didelon}, P., {et~al.} 2012, \aap, 542, A81, \dodoi{10.1051/0004-6361/201218797}

\bibitem[{{Mestel} \& {Spitzer}(1956)}]{mestel_spitzer_1956}
{Mestel}, L., \& {Spitzer}, Jr., L. 1956, \mnras, 116, 503, \dodoi{10.1093/mnras/116.5.503}

\bibitem[{{Mouschovias}(1976)}]{mouschovias_1976}
{Mouschovias}, T.~C. 1976, \apj, 207, 141, \dodoi{10.1086/154478}

\bibitem[{{Myers} \& {Benson}(1983)}]{1983_Myers}
{Myers}, P.~C., \& {Benson}, P.~J. 1983, \apj, 266, 309, \dodoi{10.1086/160780}

\bibitem[{{Ortiz-Le{\'o}n} {et~al.}(2018){Ortiz-Le{\'o}n}, {Loinard}, {Dzib}, {Kounkel}, {Galli}, {Tobin}, {Evans}, {Hartmann}, {Rodr{\'\i}guez}, {Brice{\~n}o}, {Torres}, \& {Mioduszewski}}]{2018Ortiz-Leon}
{Ortiz-Le{\'o}n}, G.~N., {Loinard}, L., {Dzib}, S.~A., {et~al.} 2018, \apjl, 869, L33, \dodoi{10.3847/2041-8213/aaf6ad}

\bibitem[{{Pandhi} {et~al.}(2023){Pandhi}, {Friesen}, {Fissel}, {Pineda}, {Caselli}, {Chen}, {Di Francesco}, {Ginsburg}, {Kirk}, {Myers}, {Offner}, {Punanova}, {Quan}, {Redaelli}, {Rosolowsky}, {Scibelli}, {Seo}, \& {Shirley}}]{Pandhi2023}
{Pandhi}, A., {Friesen}, R.~K., {Fissel}, L., {et~al.} 2023, \mnras, 525, 364, \dodoi{10.1093/mnras/stad2283}

\bibitem[{{Pattle} \& {Fissel}(2019)}]{Pattle2019}
{Pattle}, K., \& {Fissel}, L. 2019, Frontiers in Astronomy and Space Sciences, 6, 15, \dodoi{10.3389/fspas.2019.00015}

\bibitem[{{Pattle} {et~al.}(2023){Pattle}, {Fissel}, {Tahani}, {Liu}, \& {Ntormousi}}]{Pattle2023}
{Pattle}, K., {Fissel}, L., {Tahani}, M., {Liu}, T., \& {Ntormousi}, E. 2023, in Astronomical Society of the Pacific Conference Series, Vol. 534, Protostars and Planets VII, ed. S.~{Inutsuka}, Y.~{Aikawa}, T.~{Muto}, K.~{Tomida}, \& M.~{Tamura}, 193, \dodoi{10.48550/arXiv.2203.11179}

\bibitem[{{Pattle} {et~al.}(2017){Pattle}, {Ward-Thompson}, {Berry}, {Hatchell}, {Chen}, {Pon}, {Koch}, {Kwon}, {Kim}, {Bastien}, {Cho}, {Coud{\'e}}, {Di Francesco}, {Fuller}, {Furuya}, {Graves}, {Johnstone}, {Kirk}, {Kwon}, {Lee}, {Matthews}, {Mottram}, {Parsons}, {Sadavoy}, {Shinnaga}, {Soam}, {Hasegawa}, {Lai}, {Qiu}, \& {Friberg}}]{pattle_2017}
{Pattle}, K., {Ward-Thompson}, D., {Berry}, D., {et~al.} 2017, \apj, 846, 122, \dodoi{10.3847/1538-4357/aa80e5}

\bibitem[{{Pattle} {et~al.}(2021){Pattle}, {Lai}, {Di Francesco}, {Sadavoy}, {Ward-Thompson}, {Johnstone}, {Hoang}, {Arzoumanian}, {Bastien}, {Bourke}, {Coud{\'e}}, {Doi}, {Eswaraiah}, {Fanciullo}, {Furuya}, {Hwang}, {Hull}, {Kang}, {Kim}, {Kirchschlager}, {Kwon}, {Kwon}, {Lee}, {Liu}, {Redman}, {Soam}, {Tahani}, {Tamura}, \& {Tang}}]{BISTRO_L1689}
{Pattle}, K., {Lai}, S.-P., {Di Francesco}, J., {et~al.} 2021, \apj, 907, 88, \dodoi{10.3847/1538-4357/abcc6c}

\bibitem[{{Pattle} {et~al.}(2025){Pattle}, {Di Francesco}, {Hatchell}, {Kirk}, {Sadavoy}, {Ward-Thompson}, {Johnstone}, {Nittala}, {Kerr}, {Keown}, {Butner}, {Coud{\'e}}, {Currie}, {Friesen}, {Jenness}, {Knee}, \& {White}}]{2025Pattle_JCMT}
{Pattle}, K., {Di Francesco}, J., {Hatchell}, J., {et~al.} 2025, \mnras, 543, 3547, \dodoi{10.1093/mnras/staf1549}

\bibitem[{{Pezzuto} {et~al.}(2021){Pezzuto}, {Benedettini}, {Di Francesco}, {Palmeirim}, {Sadavoy}, {Schisano}, {Li Causi}, {Andr{\'e}}, {Arzoumanian}, {Bernard}, {Bontemps}, {Elia}, {Fiorellino}, {Kirk}, {K{\"o}nyves}, {Ladjelate}, {Men'shchikov}, {Motte}, {Piccotti}, {Schneider}, {Spinoglio}, {Ward-Thompson}, \& {Wilson}}]{HGBSPerseus}
{Pezzuto}, S., {Benedettini}, M., {Di Francesco}, J., {et~al.} 2021, \aap, 645, A55, \dodoi{10.1051/0004-6361/201936534}

\bibitem[{{Pillai} {et~al.}(2020){Pillai}, {Clemens}, {Reissl}, {Myers}, {Kauffmann}, {Lopez-Rodriguez}, {Alves}, {Franco}, {Henshaw}, {Menten}, {Nakamura}, {Seifried}, {Sugitani}, \& {Wiesemeyer}}]{pillai_2020}
{Pillai}, T. G.~S., {Clemens}, D.~P., {Reissl}, S., {et~al.} 2020, Nature Astronomy, 4, 1195, \dodoi{10.1038/s41550-020-1172-6}

\bibitem[{{Pineda} {et~al.}(2026){Pineda}, {Friesen}, {(co-PIs)}, {Rosolowsky}, {Chac{\'o}n-Tanarro}, {Chen}, {Di Francesco}, {Kirk}, {Punanova}, {Seo}, {Shirley}, {Ginsburg}, {Offner}, {Pandhi}, {Singh}, {Quan}, {Arce}, {Caselli}, {Choudhury}, {Goodman}, {Heitsch}, {Martin}, {Matzner}, {Myers}, {Redaelli}, {Scibelli}, \& {GAS Collaboration}}]{Pineda_2026_GAS_DR2}
{Pineda}, J.~E., {Friesen}, R.~K., {(co-PIs)}, {et~al.} 2026, \apjs, 282, 18, \dodoi{10.3847/1538-4365/ae11b1}

\bibitem[{{Planck Collaboration} {et~al.}(2015){Planck Collaboration}, {Ade}, {Aghanim}, {Alina}, {Alves}, {Armitage-Caplan}, {Arnaud}, {Arzoumanian}, {Ashdown}, {Atrio-Barandela}, {Aumont}, {Baccigalupi}, {Banday}, {Barreiro}, {Battaner}, {Benabed}, {Benoit-L{\'e}vy}, {Bernard}, {Bersanelli}, {Bielewicz}, {Bock}, {Bond}, {Borrill}, {Bouchet}, {Boulanger}, {Bracco}, {Burigana}, {Butler}, {Cardoso}, {Catalano}, {Chamballu}, {Chary}, {Chiang}, {Christensen}, {Colombi}, {Colombo}, {Combet}, {Couchot}, {Coulais}, {Crill}, {Curto}, {Cuttaia}, {Danese}, {Davies}, {Davis}, {de Bernardis}, {de Gouveia Dal Pino}, {de Rosa}, {de Zotti}, {Delabrouille}, {D{\'e}sert}, {Dickinson}, {Diego}, {Donzelli}, {Dor{\'e}}, {Douspis}, {Dunkley}, {Dupac}, {Efstathiou}, {En{\ss}lin}, {Eriksen}, {Falgarone}, {Ferri{\`e}re}, {Finelli}, {Forni}, {Frailis}, {Fraisse}, {Franceschi}, {Galeotta}, {Ganga}, {Ghosh}, {Giard}, {Giraud-H{\'e}raud}, {Gonz{\'a}lez-Nuevo}, {G{\'o}rski}, {Gregorio}, {Gruppuso}, {Guillet}, {Hansen}, {Harrison},
  {Helou}, {Hern{\'a}ndez-Monteagudo}, {Hildebrandt}, {Hivon}, {Hobson}, {Holmes}, {Hornstrup}, {Huffenberger}, {Jaffe}, {Jaffe}, {Jones}, {Juvela}, {Keih{\"a}nen}, {Keskitalo}, {Kisner}, {Kneissl}, {Knoche}, {Kunz}, {Kurki-Suonio}, {Lagache}, {L{\"a}hteenm{\"a}ki}, {Lamarre}, {Lasenby}, {Lawrence}, {Leahy}, {Leonardi}, {Levrier}, {Liguori}, {Lilje}, {Linden-V{\o}rnle}, {L{\'o}pez-Caniego}, {Lubin}, {Mac{\'\i}as-P{\'e}rez}, {Maffei}, {Magalh{\~a}es}, {Maino}, {Mandolesi}, {Maris}, {Marshall}, {Martin}, {Mart{\'\i}nez-Gonz{\'a}lez}, {Masi}, {Matarrese}, {Mazzotta}, {Melchiorri}, {Mendes}, {Mennella}, {Migliaccio}, {Miville-Desch{\^e}nes}, {Moneti}, {Montier}, {Morgante}, {Mortlock}, {Munshi}, {Murphy}, {Naselsky}, {Nati}, {Natoli}, {Netterfield}, {Noviello}, {Novikov}, {Novikov}, {Oxborrow}, {Pagano}, {Pajot}, {Paladini}, {Paoletti}, {Pasian}, {Pearson}, {Perdereau}, {Perotto}, {Perrotta}, {Piacentini}, {Piat}, {Pietrobon}, {Plaszczynski}, {Poidevin}, {Pointecouteau}, {Polenta}, {Popa}, {Pratt}, {Prunet},
  {Puget}, {Rachen}, {Reach}, {Rebolo}, {Reinecke}, {Remazeilles}, {Renault}, {Ricciardi}, {Riller}, {Ristorcelli}, {Rocha}, {Rosset}, {Roudier}, {Rubi{\~n}o-Mart{\'\i}n}, {Rusholme}, {Sandri}, {Savini}, {Scott}, {Spencer}, {Stolyarov}, {Stompor}, {Sudiwala}, {Sutton}, {Suur-Uski}, {Sygnet}, {Tauber}, {Terenzi}, {Toffolatti}, {Tomasi}, {Tristram}, {Tucci}, {Umana}, {Valenziano}, {Valiviita}, {Van Tent}, {Vielva}, {Villa}, {Wade}, {Wandelt}, {Zacchei}, \& {Zonca}}]{2015Planck}
{Planck Collaboration}, {Ade}, P.~A.~R., {Aghanim}, N., {et~al.} 2015, \aap, 576, A104, \dodoi{10.1051/0004-6361/201424082}

\bibitem[{{Planck Collaboration} {et~al.}(2016){Planck Collaboration}, {Ade}, {Aghanim}, {Alves}, {Arnaud}, {Arzoumanian}, {Ashdown}, {Aumont}, {Baccigalupi}, {Banday}, {Barreiro}, {Bartolo}, {Battaner}, {Benabed}, {Beno{\^\i}t}, {Benoit-L{\'e}vy}, {Bernard}, {Bersanelli}, {Bielewicz}, {Bock}, {Bonavera}, {Bond}, {Borrill}, {Bouchet}, {Boulanger}, {Bracco}, {Burigana}, {Calabrese}, {Cardoso}, {Catalano}, {Chiang}, {Christensen}, {Colombo}, {Combet}, {Couchot}, {Crill}, {Curto}, {Cuttaia}, {Danese}, {Davies}, {Davis}, {de Bernardis}, {de Rosa}, {de Zotti}, {Delabrouille}, {Dickinson}, {Diego}, {Dole}, {Donzelli}, {Dor{\'e}}, {Douspis}, {Ducout}, {Dupac}, {Efstathiou}, {Elsner}, {En{\ss}lin}, {Eriksen}, {Falceta-Gon{\c{c}}alves}, {Falgarone}, {Ferri{\`e}re}, {Finelli}, {Forni}, {Frailis}, {Fraisse}, {Franceschi}, {Frejsel}, {Galeotta}, {Galli}, {Ganga}, {Ghosh}, {Giard}, {Gjerl{\o}w}, {Gonz{\'a}lez-Nuevo}, {G{\'o}rski}, {Gregorio}, {Gruppuso}, {Gudmundsson}, {Guillet}, {Harrison}, {Helou}, {Hennebelle},
  {Henrot-Versill{\'e}}, {Hern{\'a}ndez-Monteagudo}, {Herranz}, {Hildebrandt}, {Hivon}, {Holmes}, {Hornstrup}, {Huffenberger}, {Hurier}, {Jaffe}, {Jaffe}, {Jones}, {Juvela}, {Keih{\"a}nen}, {Keskitalo}, {Kisner}, {Knoche}, {Kunz}, {Kurki-Suonio}, {Lagache}, {Lamarre}, {Lasenby}, {Lattanzi}, {Lawrence}, {Leonardi}, {Levrier}, {Liguori}, {Lilje}, {Linden-V{\o}rnle}, {L{\'o}pez-Caniego}, {Lubin}, {Mac{\'\i}as-P{\'e}rez}, {Maino}, {Mandolesi}, {Mangilli}, {Maris}, {Martin}, {Mart{\'\i}nez-Gonz{\'a}lez}, {Masi}, {Matarrese}, {Melchiorri}, {Mendes}, {Mennella}, {Migliaccio}, {Miville-Desch{\^e}nes}, {Moneti}, {Montier}, {Morgante}, {Mortlock}, {Munshi}, {Murphy}, {Naselsky}, {Nati}, {Netterfield}, {Noviello}, {Novikov}, {Novikov}, {Oppermann}, {Oxborrow}, {Pagano}, {Pajot}, {Paladini}, {Paoletti}, {Pasian}, {Perotto}, {Pettorino}, {Piacentini}, {Piat}, {Pierpaoli}, {Pietrobon}, {Plaszczynski}, {Pointecouteau}, {Polenta}, {Ponthieu}, {Pratt}, {Prunet}, {Puget}, {Rachen}, {Reinecke}, {Remazeilles}, {Renault},
  {Renzi}, {Ristorcelli}, {Rocha}, {Rossetti}, {Roudier}, {Rubi{\~n}o-Mart{\'\i}n}, {Rusholme}, {Sandri}, {Santos}, {Savelainen}, {Savini}, {Scott}, {Soler}, {Stolyarov}, {Sudiwala}, {Sutton}, {Suur-Uski}, {Sygnet}, {Tauber}, {Terenzi}, {Toffolatti}, {Tomasi}, {Tristram}, {Tucci}, {Umana}, {Valenziano}, {Valiviita}, {Van Tent}, {Vielva}, {Villa}, {Wade}, {Wandelt}, {Wehus}, {Ysard}, {Yvon}, \& {Zonca}}]{2016Planck}
---. 2016, \aap, 586, A138, \dodoi{10.1051/0004-6361/201525896}

\bibitem[{{Poidevin} {et~al.}(2010){Poidevin}, {Bastien}, \& {Matthews}}]{poidevin_2010}
{Poidevin}, F., {Bastien}, P., \& {Matthews}, B.~C. 2010, \apj, 716, 893, \dodoi{10.1088/0004-637X/716/2/893}

\bibitem[{{Qiu} {et~al.}(2014){Qiu}, {Zhang}, {Menten}, {Liu}, {Tang}, \& {Girart}}]{2014Qiu}
{Qiu}, K., {Zhang}, Q., {Menten}, K.~M., {et~al.} 2014, \apjl, 794, L18, \dodoi{10.1088/2041-8205/794/1/L18}

\bibitem[{{Sadavoy} {et~al.}(2018){Sadavoy}, {Myers}, {Stephens}, {Tobin}, {Kwon}, {Segura-Cox}, {Henning}, {Commer{\c{c}}on}, \& {Looney}}]{2018Sadavoy}
{Sadavoy}, S.~I., {Myers}, P.~C., {Stephens}, I.~W., {et~al.} 2018, \apj, 869, 115, \dodoi{10.3847/1538-4357/aaef81}

\bibitem[{{Sato} {et~al.}(1988){Sato}, {Tamura}, {Nagata}, {Kaifu}, {Hough}, {McLean}, {Garden}, \& {Gatley}}]{sato_1988}
{Sato}, S., {Tamura}, M., {Nagata}, T., {et~al.} 1988, \mnras, 230, 321, \dodoi{10.1093/mnras/230.2321}

\bibitem[{{Sharma} {et~al.}(2022){Sharma}, {Gopinathan}, {Soam}, {Lee}, \& {Seshadri}}]{sharma2022}
{Sharma}, E., {Gopinathan}, M., {Soam}, A., {Lee}, C.~W., \& {Seshadri}, T.~R. 2022, \mnras, 517, 1138, \dodoi{10.1093/mnras/stac2487}

\bibitem[{{Sharma} {et~al.}(2025){Sharma}, {Pattle}, {Li}, {Lee}, {Gopinathan}, {Ching}, {Tahani}, \& {Kim}}]{sharma_2025}
{Sharma}, E., {Pattle}, K., {Li}, D., {et~al.} 2025, \apj, 994, 56, \dodoi{10.3847/1538-4357/ae03a9}

\bibitem[{{Smirnov}(1948)}]{smirnov1948}
{Smirnov}, N. 1948, The Annals of Mathematical Statistics, 19, 279

\bibitem[{{Soam} {et~al.}(2018){Soam}, {Pattle}, {Ward-Thompson}, {Lee}, {Sadavoy}, {Koch}, {Kim}, {Kwon}, {Kwon}, {Arzoumanian}, {Berry}, {Hoang}, {Tamura}, {Lee}, {Liu}, {Kim}, {Johnstone}, {Nakamura}, {Lyo}, {Onaka}, {Kim}, {Furuya}, {Hasegawa}, {Lai}, {Bastien}, {Chung}, {Kim}, {Parsons}, {Rawlings}, {Mairs}, {Graves}, {Robitaille}, {Liu}, {Whitworth}, {Eswaraiah}, {Rao}, {Yoo}, {Houde}, {Kang}, {Doi}, {Choi}, {Kang}, {Coud{\'e}}, {Li}, {Matsumura}, {Matthews}, {Pon}, {Di Francesco}, {Hayashi}, {Kawabata}, {Inutsuka}, {Qiu}, {Franzmann}, {Friberg}, {Greaves}, {Kirk}, {Li}, {Shinnaga}, {van Loo}, {Aso}, {Byun}, {Chen}, {Chen}, {Chen}, {Ching}, {Cho}, {Chrysostomou}, {Drabek-Maunder}, {Eyres}, {Fiege}, {Friesen}, {Fuller}, {Gledhill}, {Griffin}, {Gu}, {Hatchell}, {Holland}, {Inoue}, {Iwasaki}, {Jeong}, {Kang}, {Kemper}, {Kim}, {Kim}, {Lacaille}, {Lee}, {Li}, {Liu}, {Liu}, {Moriarty-Schieven}, {Nakanishi}, {Ohashi}, {Peretto}, {Pyo}, {Qian}, {Retter}, {Richer}, {Rigby}, {Savini}, {Scaife}, {Tang},
  {Tomisaka}, {Wang}, {Wang}, {Yen}, {Yuan}, {Zhang}, {Zhang}, {Zhou}, {Zhu}, {Andr{\'e}}, {Dowell}, {Falle}, {Tsukamoto}, {Kanamori}, {Kataoka}, {Kobayashi}, {Nagata}, {Saito}, {Seta}, {Hwang}, {Han}, {Lee}, \& {Zenko}}]{BISTRO_OphB}
{Soam}, A., {Pattle}, K., {Ward-Thompson}, D., {et~al.} 2018, \apj, 861, 65, \dodoi{10.3847/1538-4357/aac4a6}

\bibitem[{{Soler}(2019)}]{Soler2019}
{Soler}, J.~D. 2019, \aap, 629, A96, \dodoi{10.1051/0004-6361/201935779}

\bibitem[{{Strittmatter}(1966)}]{strittmatter_1966}
{Strittmatter}, P.~A. 1966, \mnras, 132, 359, \dodoi{10.1093/mnras/132.2.359}

\bibitem[{{Tahani} {et~al.}(2018){Tahani}, {Plume}, {Brown}, \& {Kainulainen}}]{2018Tahani}
{Tahani}, M., {Plume}, R., {Brown}, J.~C., \& {Kainulainen}, J. 2018, \aap, 614, A100, \dodoi{10.1051/0004-6361/201732219}

\bibitem[{{Tahani} {et~al.}(2023){Tahani}, {Bastien}, {Furuya}, {Pattle}, {Johnstone}, {Arzoumanian}, {Doi}, {Hasegawa}, {Inutsuka}, {Coud{\'e}}, {Fissel}, {Chen}, {Poidevin}, {Sadavoy}, {Friesen}, {Koch}, {Di Francesco}, {Moriarty-Schieven}, {Chen}, {Chung}, {Eswaraiah}, {Fanciullo}, {Gledhill}, {Le Gouellec}, {Hoang}, {Hwang}, {Kang}, {Kim}, {Kirchschlager}, {Kwon}, {Lee}, {Liu}, {Onaka}, {Rawlings}, {Soam}, {Tamura}, {Tang}, {Tomisaka}, {Whitworth}, {Kwon}, {Hoang}, {Redman}, {Berry}, {Ching}, {Wang}, {Lai}, {Qiu}, {Ward-Thompson}, {Houde}, {Byun}, {Chen}, {Chen}, {Cho}, {Choi}, {Choi}, {Chrysostomou}, {Diep}, {Duan}, {Fiege}, {Franzmann}, {Friberg}, {Fuller}, {Graves}, {Greaves}, {Griffin}, {Gu}, {Han}, {Hatchell}, {Hayashi}, {Hull}, {Inoue}, {Iwasaki}, {Jeong}, {Kanamori}, {Kang}, {Kang}, {Kataoka}, {Kawabata}, {Kemper}, {Kim}, {Kim}, {Kim}, {Kim}, {Kim}, {Kirk}, {Kobayashi}, {Konyves}, {Kusune}, {Lacaille}, {Law}, {Lee}, {Lee}, {Lee}, {Lee}, {Lee}, {Li}, {Li}, {Li}, {Liu}, {Liu}, {Liu}, {de Looze},
  {Lyo}, {Mairs}, {Matsumura}, {Matthews}, {Nagata}, {Nakamura}, {Nakanishi}, {Ohashi}, {Park}, {Parsons}, {Peretto}, {Pyo}, {Qian}, {Rao}, {Retter}, {Richer}, {Rigby}, {Saito}, {Savini}, {Scaife}, {Seta}, {Shimajiri}, {Shinnaga}, {Tang}, {Tsukamoto}, {Viti}, {Wang}, {Yen}, {Yoo}, {Yuan}, {Yun}, {Zenko}, {Zhang}, {Zhang}, {Zhang}, {Zhou}, {Zhu}, {Andr{\'e}}, {Dowell}, {Eyres}, {Falle}, {van Loo}, \& {Robitaille}}]{tahani_2023}
{Tahani}, M., {Bastien}, P., {Furuya}, R.~S., {et~al.} 2023, \apj, 944, 139, \dodoi{10.3847/1538-4357/acac81}

\bibitem[{{Tritsis} {et~al.}(2025){Tritsis}, {Basu}, \& {Federrath}}]{2025Tritsis}
{Tritsis}, A., {Basu}, S., \& {Federrath}, C. 2025, \aap, 696, A35, \dodoi{10.1051/0004-6361/202453265}

\bibitem[{{Tsukamoto} {et~al.}(2023){Tsukamoto}, {Maury}, {Commercon}, {Alves}, {Cox}, {Sakai}, {Ray}, {Zhao}, \& {Machida}}]{tsukamoto_review}
{Tsukamoto}, Y., {Maury}, A., {Commercon}, B., {et~al.} 2023, in Astronomical Society of the Pacific Conference Series, Vol. 534, Protostars and Planets VII, ed. S.~{Inutsuka}, Y.~{Aikawa}, T.~{Muto}, K.~{Tomida}, \& M.~{Tamura}, 317, \dodoi{10.48550/arXiv.2209.13765}

\bibitem[{Virtanen {et~al.}(2020)Virtanen, Gommers, Oliphant, Haberland, Reddy, Cournapeau, Burovski, Peterson, Weckesser, Bright, {van der Walt}, Brett, Wilson, Millman, Mayorov, Nelson, Jones, Kern, Larson, Carey, Polat, Feng, Moore, {VanderPlas}, Laxalde, Perktold, Cimrman, Henriksen, Quintero, Harris, Archibald, Ribeiro, Pedregosa, {van Mulbregt}, \& {SciPy 1.0 Contributors}}]{2020SciPy-NMeth}
Virtanen, P., Gommers, R., Oliphant, T.~E., {et~al.} 2020, Nature Methods, 17, 261, \dodoi{10.1038/s41592-019-0686-2}

\bibitem[{{Ward-Thompson} {et~al.}(2007){Ward-Thompson}, {Di Francesco}, {Hatchell}, {Hogerheijde}, {Nutter}, {Bastien}, {Basu}, {Bonnell}, {Bowey}, {Brunt}, {Buckle}, {Butner}, {Cavanagh}, {Chrysostomou}, {Curtis}, {Davis}, {Dent}, {van Dishoeck}, {Edmunds}, {Fich}, {Fiege}, {Fissel}, {Friberg}, {Friesen}, {Frieswijk}, {Fuller}, {Gosling}, {Graves}, {Greaves}, {Helmich}, {Hills}, {Holland}, {Houde}, {Jayawardhana}, {Johnstone}, {Joncas}, {Kirk}, {Kirk}, {Knee}, {Matthews}, {Matthews}, {Matzner}, {Moriarty-Schieven}, {Naylor}, {Padman}, {Plume}, {Rawlings}, {Redman}, {Reid}, {Richer}, {Shipman}, {Simpson}, {Spaans}, {Stamatellos}, {Tsamis}, {Viti}, {Weferling}, {White}, {Whitworth}, {Wouterloot}, {Yates}, \& {Zhu}}]{JCMTGBS_WardThompson}
{Ward-Thompson}, D., {Di Francesco}, J., {Hatchell}, J., {et~al.} 2007, \pasp, 119, 855, \dodoi{10.1086/521277}

\bibitem[{{Ward-Thompson} {et~al.}(2017){Ward-Thompson}, {Pattle}, {Bastien}, {Furuya}, {Kwon}, {Lai}, {Qiu}, {Berry}, {Choi}, {Coud{\'e}}, {Di Francesco}, {Hoang}, {Franzmann}, {Friberg}, {Graves}, {Greaves}, {Houde}, {Johnstone}, {Kirk}, {Koch}, {Kwon}, {Lee}, {Li}, {Matthews}, {Mottram}, {Parsons}, {Pon}, {Rao}, {Rawlings}, {Shinnaga}, {Sadavoy}, {van Loo}, {Aso}, {Byun}, {Eswaraiah}, {Chen}, {Chen}, {Chen}, {Ching}, {Cho}, {Chrysostomou}, {Chung}, {Doi}, {Drabek-Maunder}, {Eyres}, {Fiege}, {Friesen}, {Fuller}, {Gledhill}, {Griffin}, {Gu}, {Hasegawa}, {Hatchell}, {Hayashi}, {Holland}, {Inoue}, {Inutsuka}, {Iwasaki}, {Jeong}, {Kang}, {Kang}, {Kang}, {Kawabata}, {Kemper}, {Kim}, {Kim}, {Kim}, {Kim}, {Kim}, {Kim}, {Lacaille}, {Lee}, {Lee}, {Li}, {Li}, {Liu}, {Liu}, {Liu}, {Liu}, {Lyo}, {Mairs}, {Matsumura}, {Moriarty-Schieven}, {Nakamura}, {Nakanishi}, {Ohashi}, {Onaka}, {Peretto}, {Pyo}, {Qian}, {Retter}, {Richer}, {Rigby}, {Robitaille}, {Savini}, {Scaife}, {Soam}, {Tamura}, {Tang}, {Tomisaka}, {Wang},
  {Wang}, {Whitworth}, {Yen}, {Yoo}, {Yuan}, {Zhang}, {Zhang}, {Zhou}, {Zhu}, {Andr{\'e}}, {Dowell}, {Falle}, \& {Tsukamoto}}]{2017_WardThompson}
{Ward-Thompson}, D., {Pattle}, K., {Bastien}, P., {et~al.} 2017, \apj, 842, 66, \dodoi{10.3847/1538-4357/aa70a0}

\bibitem[{{Zucker} {et~al.}(2020){Zucker}, {Speagle}, {Schlafly}, {Green}, {Finkbeiner}, {Goodman}, \& {Alves}}]{2020Zucker}
{Zucker}, C., {Speagle}, J.~S., {Schlafly}, E.~F., {et~al.} 2020, \aap, 633, A51, \dodoi{10.1051/0004-6361/201936145}

\end{thebibliography}
\bibliographystyle{aasjournal}

\appendix

\section{Cloud-scale and core-scale field figures} \label{appendix:vectors}

In this section, we present equivalents of Figure \ref{fig:ngc1333_visual_results} for each of the remaining star-forming regions.

\begin{figure*}[hbp]
    \centering
    \includegraphics[width=\textwidth]{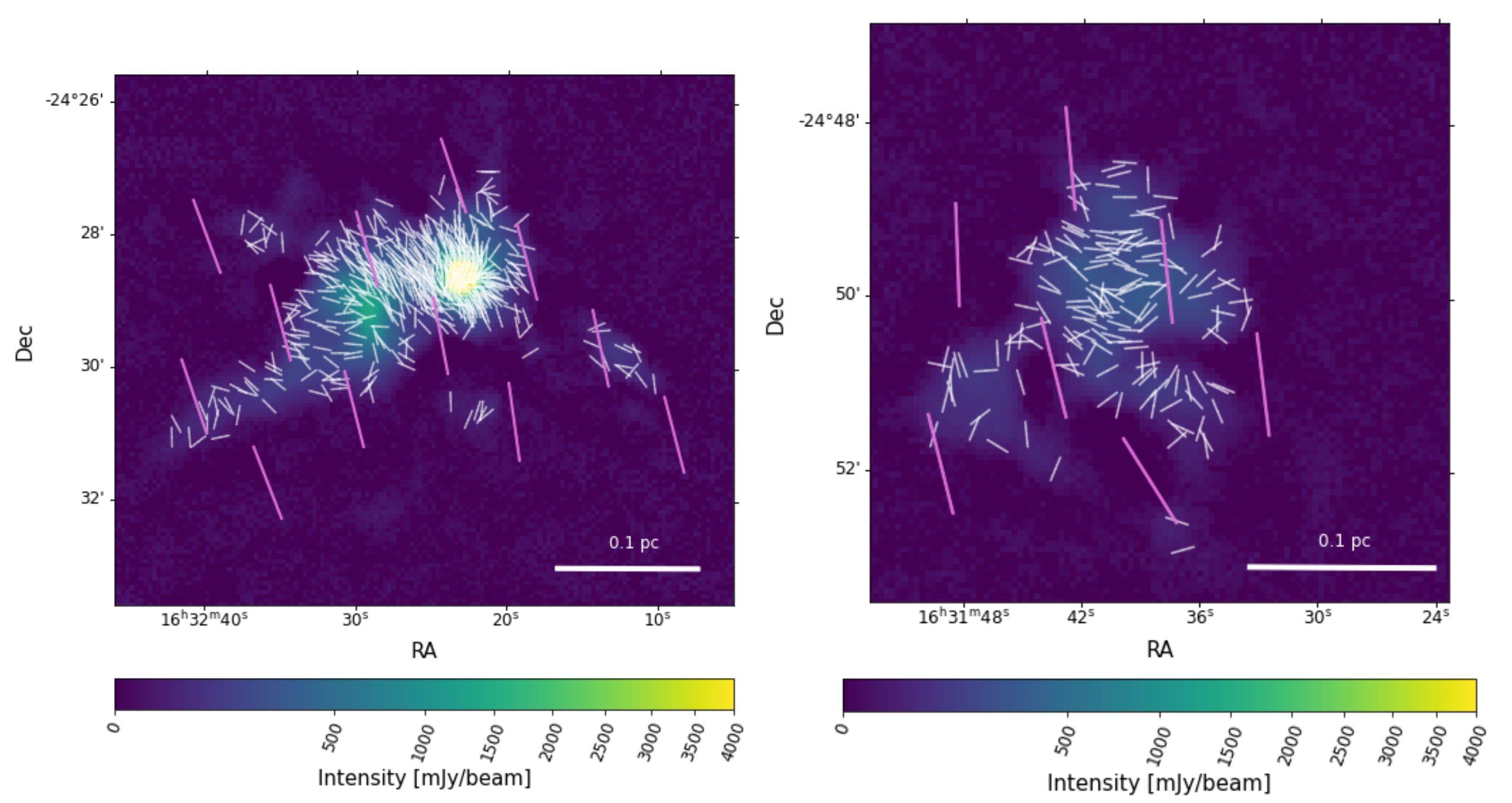}
    \caption{L1689-1 (left) and L689-2 (right) regions in the Ophiuchus molecular cloud, as in Figure \ref{fig:ngc1333_visual_results}.}
    \label{fig:l1689_visual}
\end{figure*}

\begin{figure*}
    \centering
    \includegraphics[width=\textwidth]{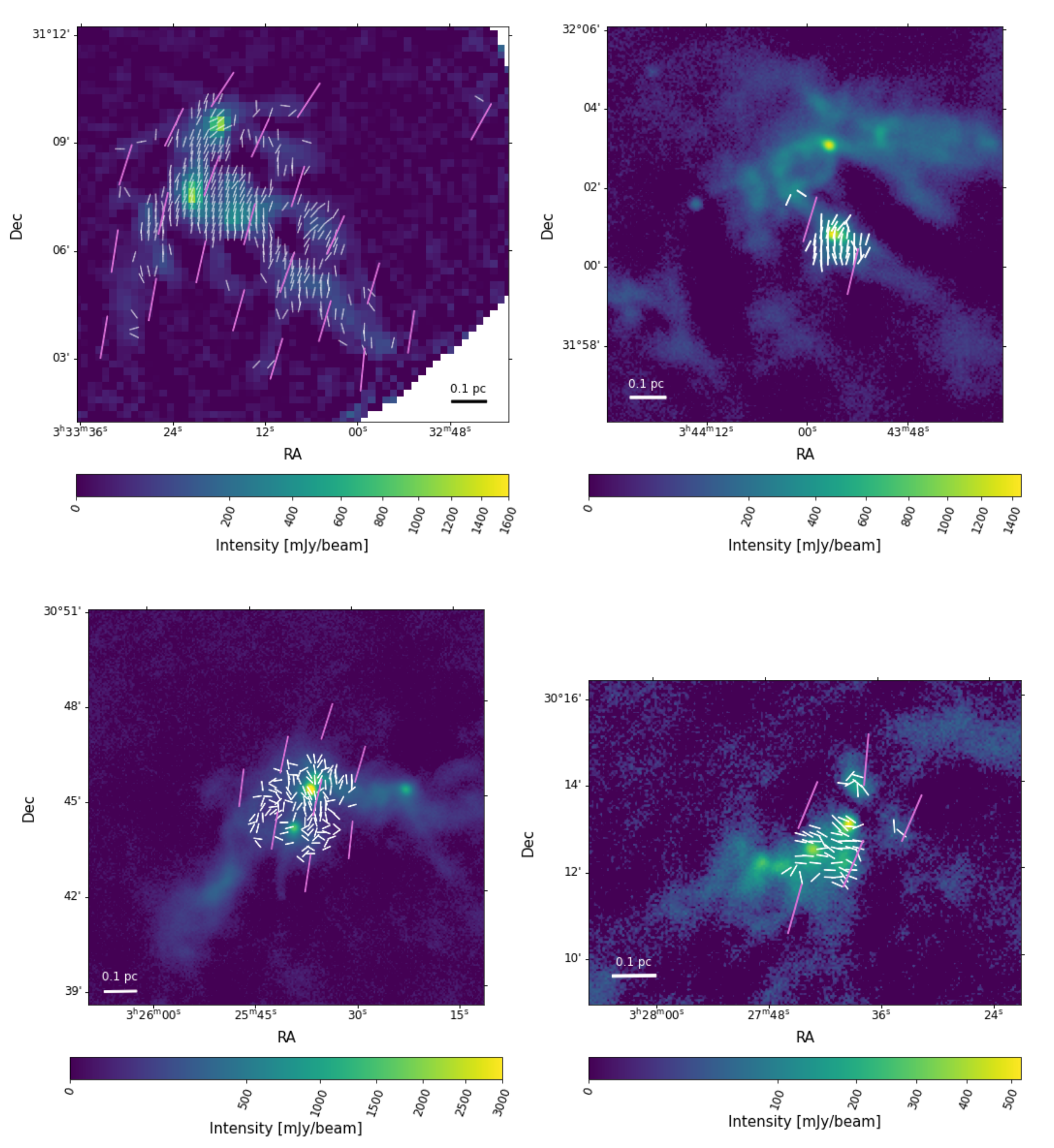}
    \caption{B1 (top left), IC348 (top right), L1448 (bottom left), and L1455 (bottom right) regions in the Perseus molecular cloud, as in Figure \ref{fig:ngc1333_visual_results}.}
    \label{fig:perseus_visual}
\end{figure*}

\begin{figure*}
    \centering
    \includegraphics[width=\textwidth]{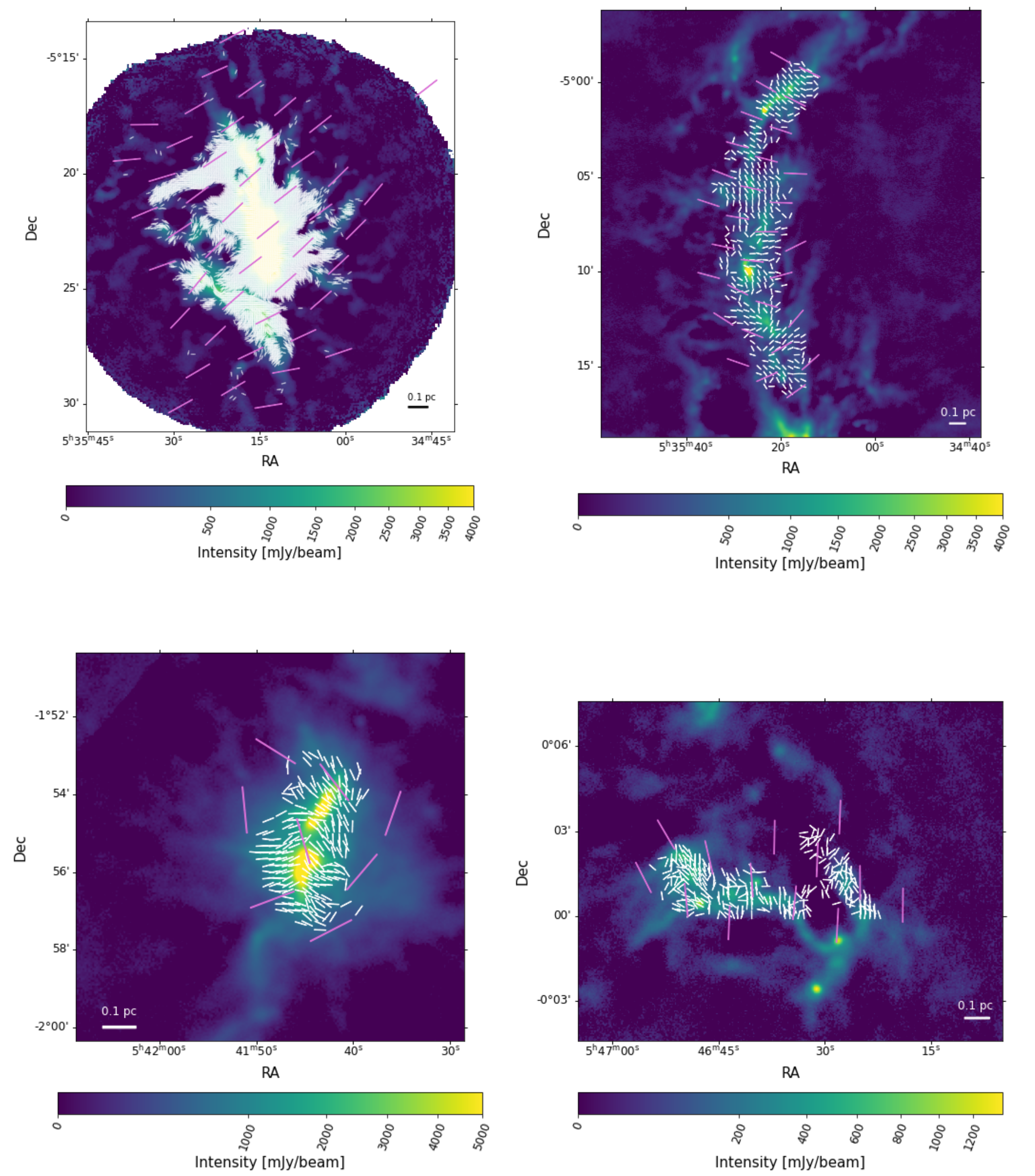}
    \caption{OMC 1 (top left), OMC 2/OMC 3 (top right), NGC 2024 (bottom left), and NGC 2068 (bottom right) regions in the Orion A and Orion B molecular clouds, as in Figure \ref{fig:ngc1333_visual_results}.}
    \label{fig:orion_visual}
\end{figure*}

\clearpage

\section{Core magnetic field orientation figures} \label{appendix:cores}

In this section, we present equivalents of Figure \ref{fig:ngc1333_core_results} for each of the remaining star-forming regions.

\begin{figure*}[hbp]
    \centering
    \includegraphics[width=\textwidth]{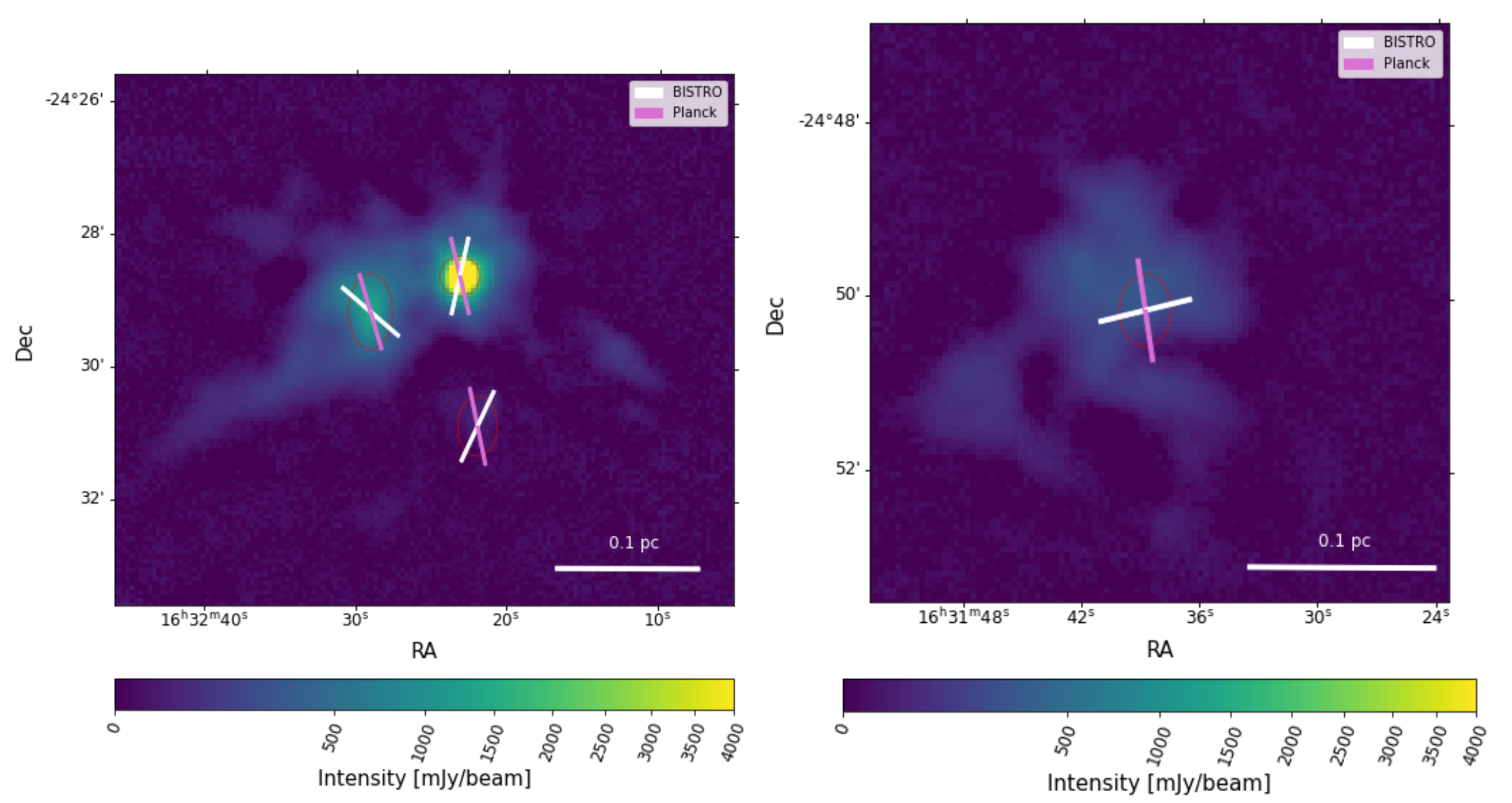}
    \caption{L1689-1 (left) and L689-2 (right) regions in the Ophiuchus molecular cloud, as in Figure \ref{fig:ngc1333_core_results}.}
    \label{fig:l1689_core}
\end{figure*}

\begin{figure*}
    \centering
    \includegraphics[width=\textwidth]{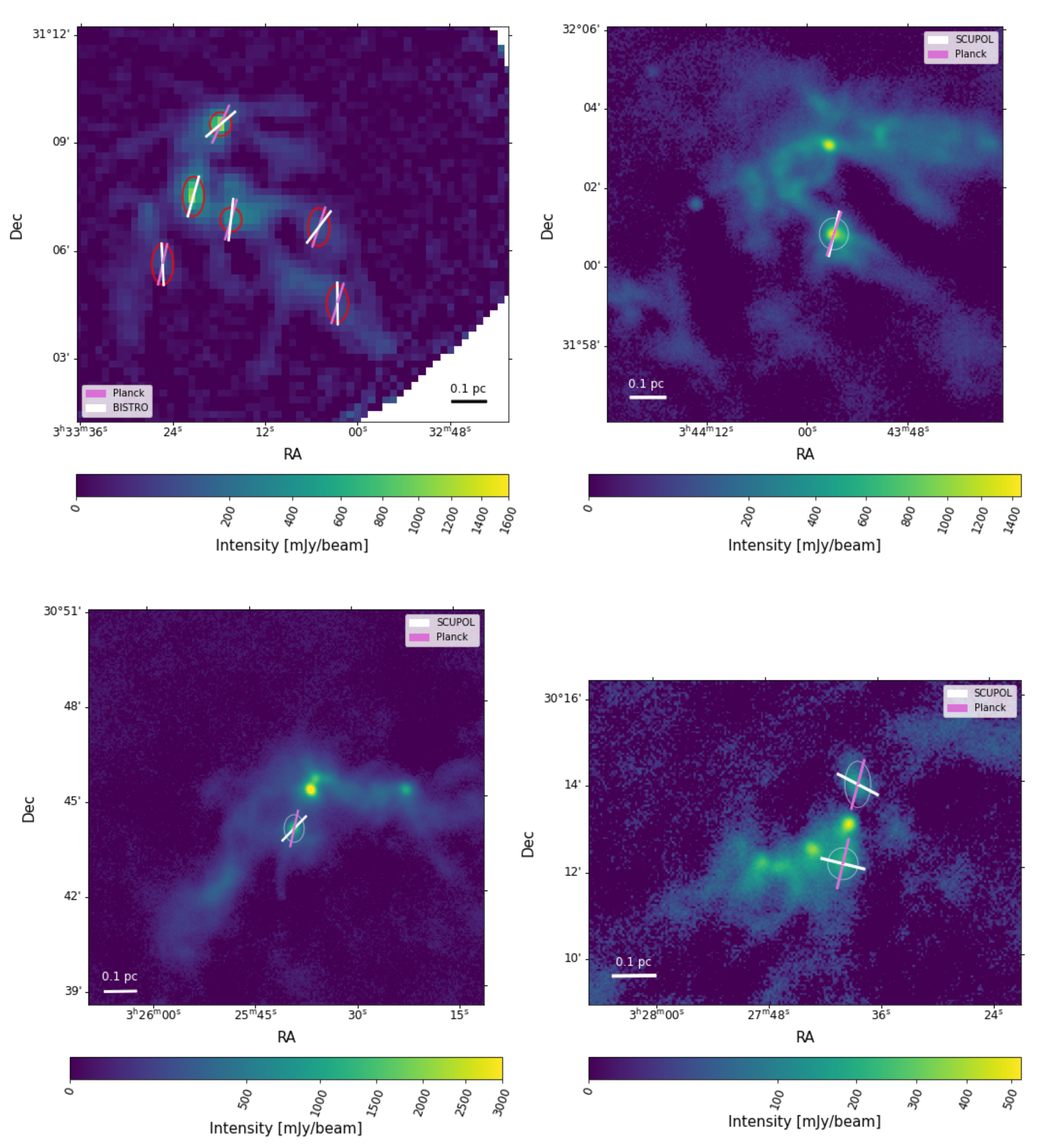}
    \caption{B1 (top left), IC348 (top right), L1448 (bottom left), and L1455 (bottom right) regions in the Perseus molecular cloud, as in Figure \ref{fig:ngc1333_core_results}.}
    \label{fig:perseus_core}
\end{figure*}

\begin{figure*}
    \centering
    \includegraphics[width=\textwidth]{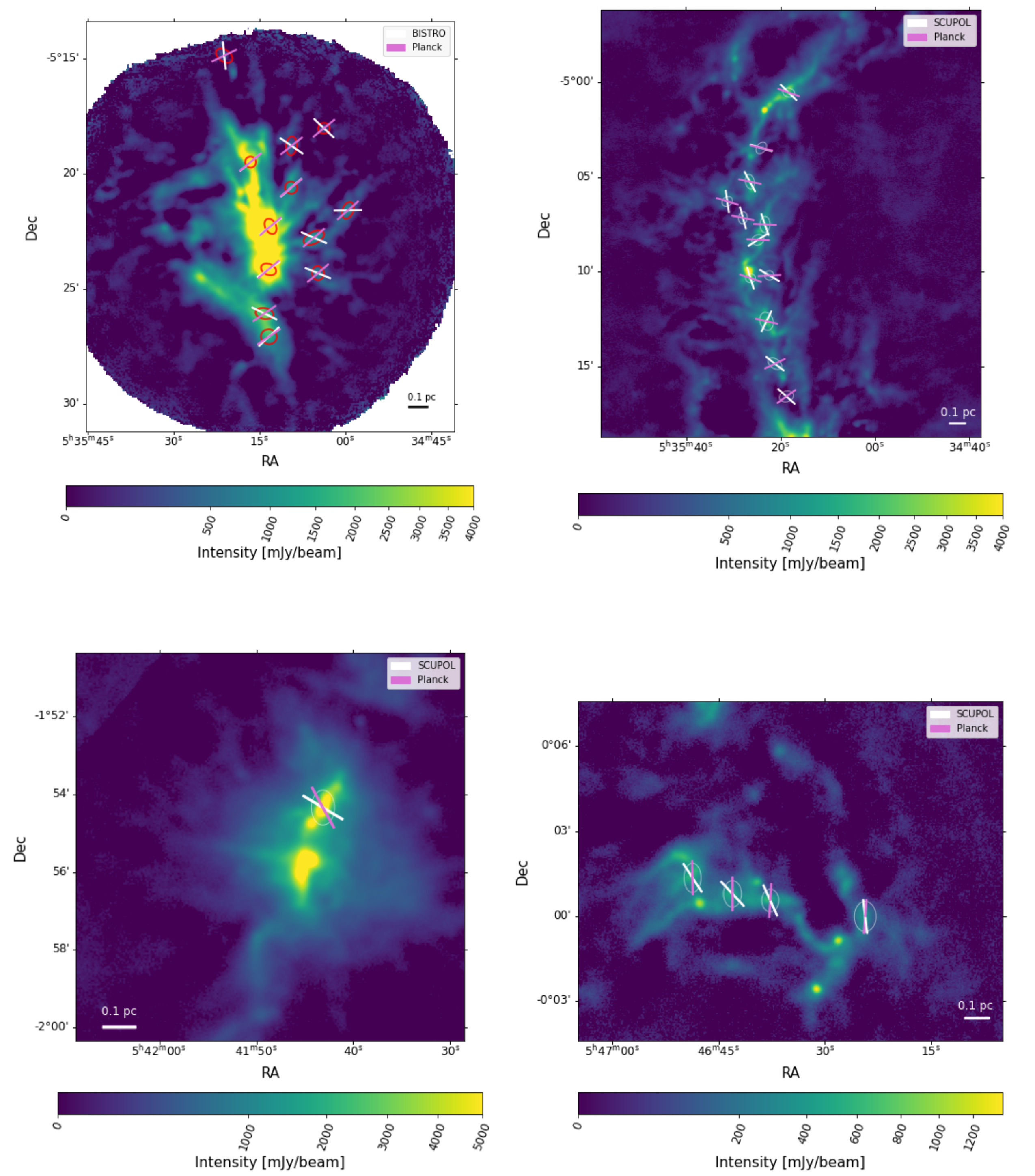}
    \caption{OMC 1 (top left), OMC 2/OMC 3 (top right), NGC 2024 (bottom left), and NGC 2068 (bottom right) regions in the Orion A and Orion B molecular clouds, as in Figure \ref{fig:ngc1333_core_results}. \label{fig:orion_core}} 
\end{figure*}

\clearpage

\section{Full core catalogue} \label{appendix:core_catalogue}

In this section, we present the full catalog of 79 cores analyzed in this paper, including information about the average core-scale magnetic field orientation $\theta_{B}$, standard deviation of the core-scale magnetic field orientations of all vectors matched to the core, number of vectors matched to the core, and alignments between $\theta_{B}$ and $\theta_\mathrm{Planck}$, $\theta_{C}$, and $\theta_{G}$.

\startlongtable
\begin{deluxetable*}{lccccccccc} 
\tabletypesize{\footnotesize} 
\tablecolumns{10} 
\tablecaption{Table containing the final core catalog produced for the 79 cores analyzed in this paper, including the source of the data, the region, the position of the core, the mean core-scale magnetic field orientation $\theta_{B}$, standard deviation of the core-scale magnetic field orientations of all vectors matched to the core, number of vectors matched to the core, and alignments between $\theta_{B}$ and $\theta_\mathrm{Planck}$, $\theta_{C}$, and $\theta_{G}$. Angles are shown in degrees, and are measured east of north in equatorial coordinates. \label{tab:core_catalogue}} 
\tablehead{ 
\colhead{Dataset \tablenotemark{a}} & \colhead{Region} & \colhead{R.A.} & \colhead{Dec.} & \colhead{$\theta_{B}$} & \colhead{$\sigma_B$} & \colhead{\# of Vectors \tablenotemark{b}} & \colhead{$|\theta_{B} - \theta_\mathrm{Planck}|$} & \colhead{$|\theta_{B} - \theta_{C}|$} & \colhead{$|\theta_{B} - \theta_{G}|$} \\ 
\colhead{} & \colhead{} & \colhead{} & \colhead{} & \colhead{($^{\circ}$)} & \colhead{($^{\circ}$)} & \colhead{} & \colhead{($^{\circ}$)} & \colhead{($^{\circ}$)} & \colhead{($^{\circ}$)}
} 
\startdata 
 BISTRO &    NGC 1333 &  52.1650 &  31.3068 &  51 &          62 &            20 &                   86 &                       54 &                              56 \\
 BISTRO &    NGC 1333 &  52.1963 &  31.3542 &  68 &           2 &             2 &                   80 &                       72 &                              12 \\
 BISTRO &    NGC 1333 &  52.1964 &  31.2532 & 114 &           8 &             4 &                   35 &                       41 &                               6 \\
 BISTRO &    NGC 1333 &  52.2129 &  31.3040 & 163 &          26 &             2 &                   17 &                       59 &                              76 \\
 BISTRO &    NGC 1333 &  52.2312 &  31.2433 &  70 &          14 &            16 &                   84 &                        4 &                              66 \\
 BISTRO &    NGC 1333 &  52.2501 &  31.3592 & 111 &          25 &            27 &                   40 &                       41 &                              16 \\
 BISTRO &    NGC 1333 &  52.2521 &  31.1998 &  66 &          19 &            14 &                   89 &                       47 &                              13 \\
 BISTRO &    NGC 1333 &  52.2557 &  31.3403 & 115 &          16 &            47 &                   33 &                       12 &                              79 \\
 BISTRO &    NGC 1333 &  52.2581 &  31.2600 & 162 &          10 &            27 &                    6 &                       31 &                               3 \\
 BISTRO &    NGC 1333 &  52.2705 &  31.3123 & 115 &          13 &            48 &                   32 &                       12 &                              34 \\
 BISTRO &    NGC 1333 &  52.2799 &  31.2895 & 124 &          17 &            38 &                   23 &                       72 &                              19 \\
 BISTRO &    NGC 1333 &  52.2874 &  31.2548 &   2 &          31 &            23 &                   28 &                       52 &                              61 \\
 BISTRO &    NGC 1333 &  52.2970 &  31.3083 & 108 &          34 &            30 &                   39 &                       26 &                              50 \\
 BISTRO &    NGC 1333 &  52.3159 &  31.3421 & 140 &          25 &             7 &                    2 &                       52 &                              25 \\
 BISTRO &    NGC 1333 &  52.3266 &  31.4187 & 164 &          36 &             7 &                   17 &                       76 &                              22 \\
 BISTRO &    NGC 1333 &  52.3288 &  31.3871 & 101 &          13 &             3 &                   47 &                        1 &                              43 \\
 BISTRO &          B1 &  53.2608 &  31.0756 &   1 &          13 &             6 &                   20 &                        3 &                              32 \\
 BISTRO &          B1 &  53.2708 &  31.1110 & 138 &          12 &            14 &                   20 &                       90 &                              77 \\
 BISTRO &          B1 &  53.3185 &  31.1145 & 173 &          11 &            16 &                   12 &                       78 &                              30 \\
 BISTRO &          B1 &  53.3240 &  31.1588 & 126 &          19 &             9 &                   26 &                       26 &                              61 \\
 BISTRO &          B1 &  53.3388 &  31.1252 & 162 &           5 &            21 &                    1 &                       69 &                              73 \\
 BISTRO &          B1 &  53.3555 &  31.0938 &   3 &          11 &             5 &                   18 &                       68 &                              17 \\
 BISTRO &       OMC 1 &  83.8060 &  -5.4023 & 124 &          23 &            73 &                    6 &                       58 &                              13 \\
 BISTRO &       OMC 1 &  83.8191 &  -5.3247 & 130 &          17 &            56 &                    1 &                       63 &                              35 \\
 BISTRO &       OMC 1 &  83.8045 &  -5.3711 & 127 &          10 &            76 &                    3 &                       79 &                              69 \\
 BISTRO &       OMC 1 &  83.7895 &  -5.3431 & 131 &           7 &            52 &                    2 &                       85 &                              74 \\
 BISTRO &       OMC 1 &  83.8090 &  -5.4343 &  64 &          36 &            76 &                   60 &                       16 &                              34 \\
 BISTRO &       OMC 1 &  83.7730 &  -5.3794 &  66 &          12 &            90 &                   64 &                       50 &                               8 \\
 BISTRO &       OMC 1 &  83.7483 &  -5.3596 &  91 &          11 &            31 &                   40 &                       68 &                              55 \\
 BISTRO &       OMC 1 &  83.8382 &  -5.2476 &   6 &           0 &             1 &                   70 &                       48 &                              34 \\
 BISTRO &       OMC 1 &  83.7701 &  -5.4050 &  68 &          20 &            15 &                   63 &                       61 &                              50 \\
 BISTRO &       OMC 1 &  83.8057 &  -5.4511 & 133 &          50 &            64 &                    9 &                       51 &                              74 \\
 BISTRO &       OMC 1 &  83.7893 &  -5.3129 &  56 &          23 &             8 &                   73 &                       61 &                              68 \\
 BISTRO &       OMC 1 &  83.7657 &  -5.3001 &  47 &           0 &             1 &                   82 &                       56 &                               4 \\
 BISTRO & L1688/L1689 & 246.5316 & -24.3420 & 140 &          44 &             6 &                   30 &                       23 &                              36 \\
 BISTRO & L1688/L1689 & 246.5425 & -24.3870 & 134 &          50 &             8 &                   44 &                       84 &                              27 \\
 BISTRO & L1688/L1689 & 246.5607 & -24.4197 & 143 &          61 &            15 &                   45 &                        5 &                              24 \\
 BISTRO & L1688/L1689 & 246.5722 & -24.3962 & 161 &          37 &             8 &                   27 &                       60 &                              41 \\
 BISTRO & L1688/L1689 & 246.6097 & -24.3738 & 101 &          30 &            73 &                   81 &                       46 &                               6 \\
 BISTRO & L1688/L1689 & 246.6152 & -24.3998 &  53 &          18 &           139 &                   24 &                       66 &                              10 \\
 BISTRO & L1688/L1689 & 246.6372 & -24.4374 & 126 &          37 &            22 &                   81 &                       57 &                              71 \\
 BISTRO & L1688/L1689 & 246.6795 & -24.4369 &  78 &          44 &             6 &                   33 &                       75 &                              82 \\
 BISTRO & L1688/L1689 & 246.6862 & -24.5547 &  50 &          33 &             5 &                   55 &                       73 &                              62 \\
 BISTRO & L1688/L1689 & 246.7059 & -24.4488 & 116 &          42 &             3 &                   10 &                       59 &                              89 \\
 BISTRO & L1688/L1689 & 246.7348 & -24.6109 & 144 &          69 &            19 &                   54 &                       90 &                              65 \\
 BISTRO & L1688/L1689 & 246.7416 & -24.5270 &  89 &          34 &             7 &                   86 &                       19 &                              54 \\
 BISTRO & L1688/L1689 & 246.7455 & -24.5729 &  58 &          46 &            66 &                   47 &                       69 &                              40 \\
 BISTRO & L1688/L1689 & 246.7716 & -24.6549 & 148 &          44 &            11 &                   56 &                       75 &                              69 \\
 BISTRO & L1688/L1689 & 246.7947 & -24.6570 & 143 &          57 &             8 &                   61 &                       37 &                              42 \\
 BISTRO & L1688/L1689 & 246.7998 & -24.4335 &  91 &          36 &             3 &                   85 &                       25 &                              89 \\
 BISTRO & L1688/L1689 & 246.8023 & -24.4969 & 136 &          32 &            26 &                   26 &                       10 &                              73 \\
 BISTRO & L1688/L1689 & 246.8210 & -24.4621 & 103 &          46 &            13 &                   65 &                        4 &                              49 \\
 BISTRO & L1688/L1689 & 246.8403 & -24.4132 & 117 &          37 &             6 &                   84 &                       12 &                              36 \\
 BISTRO & L1688/L1689 & 246.8541 & -24.4524 &  82 &          30 &            27 &                   84 &                        9 &                              35 \\
 BISTRO & L1688/L1689 & 246.8872 & -24.4419 &  34 &          31 &            20 &                   52 &                       18 &                              35 \\
 BISTRO & L1688/L1689 & 247.9118 & -24.8358 & 103 &          25 &            15 &                   87 &                       22 &                               2 \\
 BISTRO & L1688/L1689 & 248.0915 & -24.5144 & 153 &          15 &             5 &                   40 &                       58 &                              56 \\
 BISTRO & L1688/L1689 & 248.0965 & -24.4768 & 167 &          38 &            73 &                   28 &                        9 &                              70 \\
 BISTRO & L1688/L1689 & 248.1211 & -24.4858 &  52 &          27 &            36 &                   34 &                       50 &                              35 \\
 SCUPOL &       IC348 &  55.9865 &  32.0140 & 165 &          16 &            13 &                    6 &                       53 &                               7 \\
 SCUPOL &       L1448 &  51.4119 &  30.7344 & 130 &          31 &             2 &                   34 &                       57 &                              76 \\
 SCUPOL &       L1455 &  51.9092 &  30.2327 &  65 &          54 &             6 &                   83 &                        9 &                              41 \\
 SCUPOL &       L1455 &  51.9162 &  30.2022 &  77 &          19 &             7 &                   87 &                       80 &                              79 \\
 SCUPOL &    NGC 2024 &  85.4297 &  -1.9057 &  59 &          17 &            10 &                   30 &                       83 &                              82 \\
 SCUPOL &    NGC 2068 &  86.6012 &  -0.0002 &   7 &          21 &            19 &                   11 &                       19 &                              59 \\
 SCUPOL &    NGC 2068 &  86.6570 &   0.0092 &  24 &          18 &             5 &                   28 &                       82 &                              36 \\
 SCUPOL &    NGC 2068 &  86.6793 &   0.0131 &  43 &          45 &             9 &                   43 &                       37 &                              37 \\
 SCUPOL &    NGC 2068 &  86.7028 &   0.0226 &  33 &          12 &            13 &                   32 &                       77 &                               6 \\
 SCUPOL & OMC 2/OMC 3 &  83.8598 &  -5.1724 &  18 &           4 &             2 &                   53 &                       79 &                              18 \\
 SCUPOL & OMC 2/OMC 3 &  83.8603 &  -5.0876 &  28 &           4 &             4 &                   50 &                       19 &                              59 \\
 SCUPOL & OMC 2/OMC 3 &  83.8261 &  -5.0092 &  45 &           7 &             3 &                   25 &                       28 &                              14 \\
 SCUPOL & OMC 2/OMC 3 &  83.8456 &  -5.2102 & 154 &          15 &             3 &                   77 &                       42 &                               2 \\
 SCUPOL & OMC 2/OMC 3 &  83.8470 &  -5.1252 &  19 &          20 &             2 &                   69 &                       22 &                               5 \\
 SCUPOL & OMC 2/OMC 3 &  83.8532 &  -5.1387 & 122 &          25 &             2 &                   34 &                        8 &                              27 \\
 SCUPOL & OMC 2/OMC 3 &  83.8431 &  -5.1700 &  60 &          29 &             2 &                   33 &                       27 &                               4 \\
 SCUPOL & OMC 2/OMC 3 &  83.8664 &  -5.1194 &  15 &          23 &             3 &                   63 &                       21 &                              69 \\
 SCUPOL & OMC 2/OMC 3 &  83.8281 &  -5.2764 &  49 &           5 &             2 &                   77 &                       57 &                              26 \\
 SCUPOL & OMC 2/OMC 3 &  83.8502 &  -5.0579 &  73 &          54 &             3 &                    4 &                       69 &                              33 \\
 SCUPOL & OMC 2/OMC 3 &  83.8802 &  -5.1049 &   9 &          16 &             2 &                   65 &                       63 &                               8 \\
\enddata 
\tablenotetext{a}{Denotes whether JCMT data is from BISTRO or SCUPOL}
\tablenotetext{b}{Total number of magnetic field vectors matched to core}
\end{deluxetable*}

\end{document}